 \documentclass[twocolumn,showpacs,amsmath,prd,amssymb,showkeys,superscriptaddress]{revtex4-2}

\pdfoutput=1

\usepackage{graphicx}
\usepackage{dcolumn}
\usepackage{multirow}
\usepackage{mathtools}
\usepackage{scalerel}
\usepackage{bm}
\usepackage{upgreek}
\usepackage{mathrsfs}  
\usepackage[unicode=true,pdfusetitle,
 bookmarks=true,bookmarksnumbered=false,bookmarksopen=false,
 breaklinks=false,pdfborder={0 0 1},backref=false,colorlinks=true, citecolor=blue]
{hyperref}

\def\B0{{\rm B_0}}
\def\maxB0{{\rm B}_0^{max}}
\def\S0{{\rm S_0}}
\def\BI{{\rm BI}}

\def\BI0{{\rm BI_0}}

\def\Scont{ {\rm S}_0^{cont} }
\def\SPoi{ {\rm S}_0^{Poi} }
\def\dB{\sigma_{B}}

\def\RoI{{\rm RoI}}
\def\Poi{{\rm Poi}}
\def\CPoi{{\rm CPoi}}
\def\effRoI{\varepsilon_{\rm RoI}}

\def\aves0{{\rm < S_0 >}}

\def\Pgk{{\rm P}_{g}^{k \sigma}}
\def\P503s{{\rm P_{50}^{3\sigma}}}
\def\xP505s{{\rm P_{50}^{5\sigma}}}
\def\Prob903s{{\rm P_{90}^{3\sigma}}}

\def\E0{{\rm E_0}}
\def\sigmaE0{\sigma_{\rm E_0}}
\def\DeltaE0{\Delta_{\rm E_0}}
\def\Nsigma{\rm N_{\sigma}}
\def\Noptsigma{{\rm N}^{opt}_{\sigma}}
\def\Sopt{{\rm S}_0^{opt}}
\def\nobs{n_{obs}}

\def\q0{q_{0}}
\def\Hnull{ H_0 }
\def\H1{ H_1 }
\def\talpha{t_{\alpha}}

\def\nll-Pq0H0{P(q_{0}|H_{0})}
\def\alt-Pq0H1{P(q_{0}|H_{1})}

\def\mbb{ \langle m_{\beta \beta} \rangle}

\def\thalf0nu{\uptau^{0 \nu}_{\scaleto{1/2}{5pt}}}
\def\t2bb{\uptau^{2 \nu}_{\scaleto{1/2}{5pt}}}

\def\0nubb{0 \nu \beta \beta}
\def\2nubb{2 \nu \beta \beta}

\def\Qbb{ Q_{\beta \beta} }
\def\Abb{ A_{\beta \beta} }
\def\ME2{| M^{0 \nu} | ^2}
\def\tyr{{\rm ton\mbox{-}yr}}
\def\FWHM{\rm FHWM} 
\def\DeltaQbb{\Delta_{\Qbb}}

\def\BIunit{{\rm counts}/( \FWHM \mbox{-} \tyr )}

\def\ca48{\rm ^{48}Ca}
\def\ge76{\rm ^{76}Ge}
\def\se82{\rm ^{82}Se}
\def\zr96{\rm ^{96}Zr}
\def\mo100{\rm ^{100}Mo}
\def\cd116{\rm ^{116}Cd}
\def\te130{\rm ^{130}Te}
\def\xe136{\rm ^{136}Xe}
\def\nd150{\rm ^{150}Nd}

\begin{document}


\title{
Projections of Discovery Potentials 
from Expected Background
}

%

\newcommand{\as}{Institute of Physics, Academia Sinica,
Taipei 11529, Taiwan.}
\newcommand{\bhu}{Department of Physics, Banaras Hindu University,
Varanasi 221005, India.}
\newcommand{\cusb}{Department of Physics,
Central University of South Bihar, Gaya 824236, India.}
\newcommand{\hnbu}{
Department of Physics, 
H.N.B. Garhwal University, Srinagar 246174, India.
}

\newcommand{\corrhtw}{htwong@phys.sinica.edu.tw}
\newcommand{\corrlhb}{lihb@gate.sinica.edu.tw}
\newcommand{\corrms}{manu@gate.sinica.edu.tw}

\author{ M.K.~Singh } \altaffiliation[Corresponding Author: ]{ \corrms } \affiliation{ \as } \affiliation{ \bhu }
\author{ H.B.~Li }  \altaffiliation[Corresponding Author: ]{ \corrlhb } \affiliation{ \as }
\author{ H.T.~Wong } \altaffiliation[Corresponding Author: ]{ \corrhtw } \affiliation{ \as }
\author{ V.~Sharma }  \affiliation{ \as } \affiliation{ \hnbu }
\author{ L.~Singh }  \affiliation{ \as } \affiliation{ \cusb }

\collaboration{TEXONO Collaboration}



\date{\today}

\begin{abstract}

{
Background channels with their expected strength
and uncertainty levels are usually known
} 
in the searches of novel phenomena prior to the experiments 
are conducted at their design stage.
We quantitatively study the projected sensitivities in 
terms of discovery potentials. 
These are essential for the optimizations of the 
experimental specifications as well as 
of the cost-effectiveness in various investment.
Sensitivities in counting analysis are derived
with complete Poisson statistics and its continuous approximation,
and are compared with those using maximum likelihood analysis in which
additional measurables are included as signatures.
The roles and effects due to 
uncertainties in the background estimates are studied.
Two expected features to establish positive effects
are verified and quantified: 
(i) In counting-only experiments,
the required signal strength can be derived with 
complete Poisson analysis, and
the continuous approximation would underestimate 
the results. 
(ii) Incorporating continuous variables as additional constraints  
would reduce the required signal strength 
relative to that of counting-only analysis.
The formulations are applied to the case on
the experimental searches of
neutrinoless double beta decay in which
both ambient and two-neutrino background are considered.

\end{abstract}


\maketitle


\section{Introduction}


In experimental searches of new but rare phenomena, 
some knowledge of the background  is usually known
prior to the experiments.
A universal issue is then to make projections 
of the sensitivities, either in terms of
signal discovery potentials or 
as exclusion limits,
under certain statistical criteria the experimenters set
$-$ at the design stage {\it before} the experiments
are performed.

The answers to these questions would define how much exposure 
(target size times data taking time) 
would be required to achieve certain specified
sensitivities given the expected level of background.
This translates directly to the investment in 
hardware and time and manpower, 
the precise knowledge of which is getting increasingly important 
with more and more elaborate experimental projects.
The cost-effectiveness to deliver certain 
scientific goals should be known and compared 
at the proposal stage,
which can be a decade or longer before the actual data taking.

A similar but non-identical problem was addressed in
the classic paper of Ref.~\cite{feldman-cousins}.
The ``confidence interval'' results from that work
represent the knowledge of parameters {\it after} the
measurements are performed when the expected background
is known. The procedures were further refined~\cite{JJ-Gomez} with the
introduction of fluctuations to the actual background
in one particular measurement.
This work complements and expands these by considering
the projected sensitivities  prior to the measurements, such
that the statistical fluctuations of both signals 
and backgrounds have to be taken into account.

This article serves to address key aspects of this problem.
Counting analysis 
based on Poisson statistics
are described in Section~\ref{sect::poisson-formulation}.
Results are compared with those
from previous work in the literature using a continuous 
approximation~\cite{ijdeaor2016212,mbbprob:3,Singh:2018lan,Continuous_Poisson:1,
Continuous_Poisson:2,Continuous_Poisson:3,RevModPhys.95.025002,PhysRevD.106.032012}.
Additional measurable information 
such as energy are usually available. 
These are incorporated into the analysis
with the Maximum Likelihood Ratio 
method~\cite{loglikelihoodratio:1, loglikelihoodratio:2, loglikelihoodratio:3}.
The procedures and results are discussed
in Section~\ref{sect::maxlikelihood}.
The consequences of having uncertainties in
the background predictions are addressed
in Section~\ref{sect::delta-bkg}.

While the methodology and results of this work are with general validity
to many research subjects, they follow from our earlier
``counting-only'' analysis of 
the relation between background and exposure
in future neutrinoless 
double beta decay ($\0nubb$) projects~\cite{TEXONO:PRD:2020}.
Positive $\0nubb$ signals manifest as peaks in 
the measurable energy spectra at known resolution,
providing additional constraints which 
enhance the sensitivities
beyond those from simple counting methods.
Section~\ref{sect::0nubb} illustrates how
the statistical methods developed in this work
can be applied to $\0nubb$ experiments in practice. 
Detailed implications and comparison of the expected
sensitivities to the various future double beta decay projects
on different candidate isotopes under different experimental parameters
are beyond the scope of this work. 
These will be the themes of our subsequent studies,
based on the methodology developed in this work.



\section{Poisson Counting Analysis}
\label{sect::counting_analysis_poisson}

\subsection{Complete Poisson Distribution $-$ Formulation}
\label{sect::poisson-formulation}

In experimental measurements of rare events,
Poisson statistics~\cite{Workman:2022ynf}
quantifies the probability of
observing $\nobs$-events in a certain trial
given a known mean $\mu$:
\begin{equation}
\Poi (\nobs  ; \mu ) ~ =  ~
\frac{ \mu^{\nobs} \cdot e^{\mbox{-} \mu} }{ \nobs !} ~ ; ~
\nobs = 0,1,2,3.....  ~ ; ~  \mu > 0   ~~ .
\label{eq::Poisson-Complete}
\end{equation}
The Cumulative Poisson Distribution
\begin{equation}
\CPoi ( {\leq} C ; \mu ) = \sum_{i=0}^{C} \Poi ( i ; \mu ) 
\label{eq::Poisson-Summation}
\end{equation}
describes the probability of making an observation of 
an integer $C$-counts or less.
These offer a complete description, incorporating 
the discreteness of the problem and the inevitable
fluctuations among individual trials.

We denote $\B0$ as
expected average background counts
within certain Region of Interest (RoI), 
in which the signal efficiency
is denoted by $\effRoI$. 
{
In a counting-only analysis, 
the only available information
is $\nobs$, the observed number of events (``counts''). 
The selection of an RoI is not necessary, 
such that $\effRoI {\equiv} 1$. 
}
The background $\B0$ and its
uncertainty can, in principle, be predicted
with good accuracies prior to the experiments.

{ 
The sensitivity goals as discovery potentials 
for making positive observations in experiments 
are described by a set of criteria 
denoted by $\Pgk$,
under which there are two requirements to satisfy:
(i) An experimental measurement would have
certain statistical ``$p$-value'' of significance in the interval 
$[ {+} k \sigma , {+} \infty ]$ 
where $\sigma$ is the root-mean-square (RMS) of
the background-only Gaussian distribution. 
(ii) This condition is satisfied by 
a fraction $g$ of repeated identical experiments.
} 
We note that a typical choice in the 
literature~\cite{mbbprob:3,Continuous_Poisson:1,
Continuous_Poisson:2,Continuous_Poisson:3,RevModPhys.95.025002,PhysRevD.106.032012}
is with the two-sided ${\pm} 3 \sigma$ interval at 
$g {=} 50\%$ probability.
In our applications to experimental searches of rare
signals {\it in excess of} certain background, 
the selection of having one-sided interval of
${>} {+} k \sigma$ is appropriate.
{
The pre-defined discovery potential criteria 
of this study, denoted by $\P503s$, 
corresponds to the requirements of
having $g {=} 50\%$ cases with $``{>} {+} 3 \sigma$ excess'' 
$-$ that is, $p {=} 0.00135$, evaluated from the integration
of the interval $[ {+} 3 \sigma , {+} \infty ]$ in a Gaussian distribution.
}


\begin{figure}
\includegraphics[width=8.2cm]{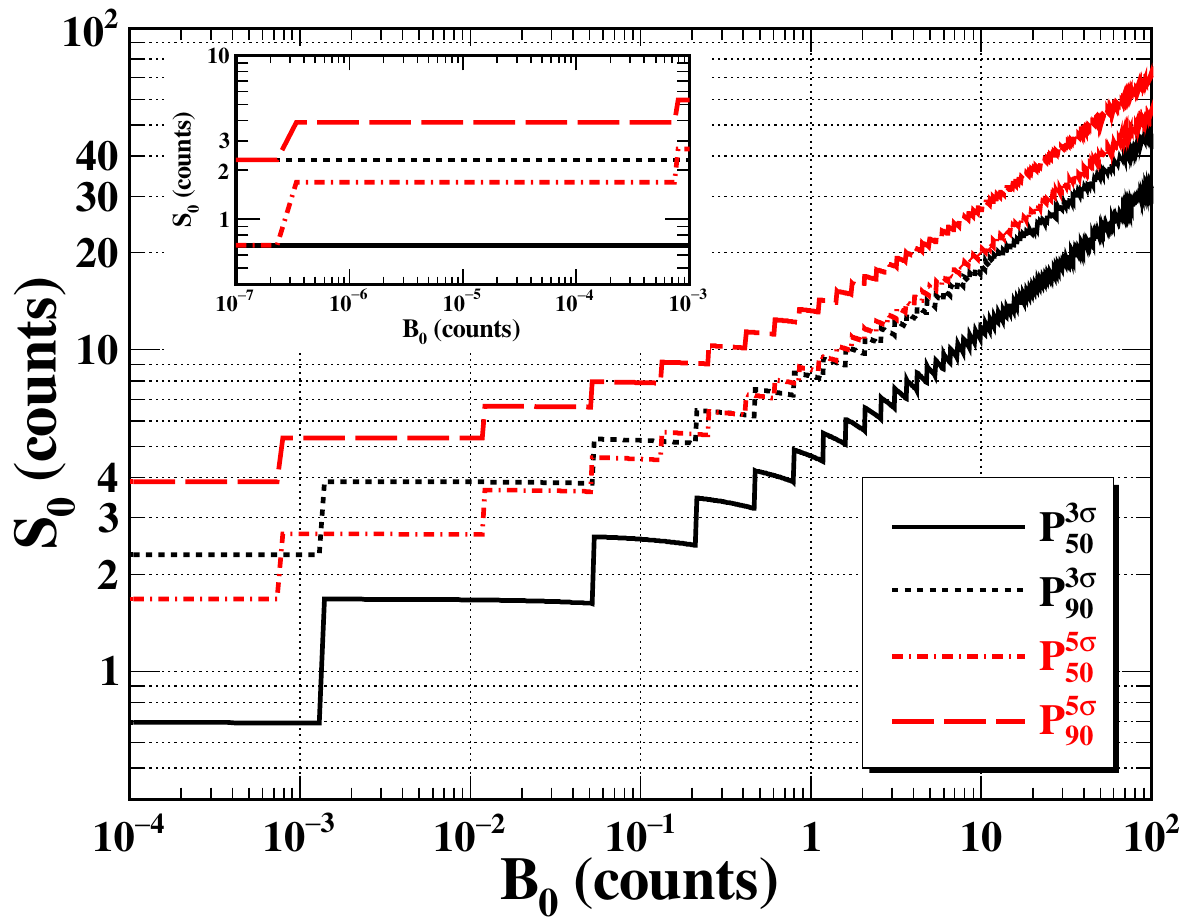}
\caption{
{
The variations of
$\S0$ versus $\B0$ in
discovery potential in counting experiments 
under the criteria $\Pgk$,
for $k {=} 3,5$ and $g {=} 50,90\%$.
The inset displays contours at $\B0 {<} 10^{\mbox{-}3}$.
The first steps at lowest $\B0$
correspond  to the transition
where an increase of $\nobs$ from 1 to 2 events
is required to positively establish the signals.
}
}
\label{fig::DP-S0vsB0}
\end{figure}


\begin{figure}[hbt]
\includegraphics[width=8.2cm]{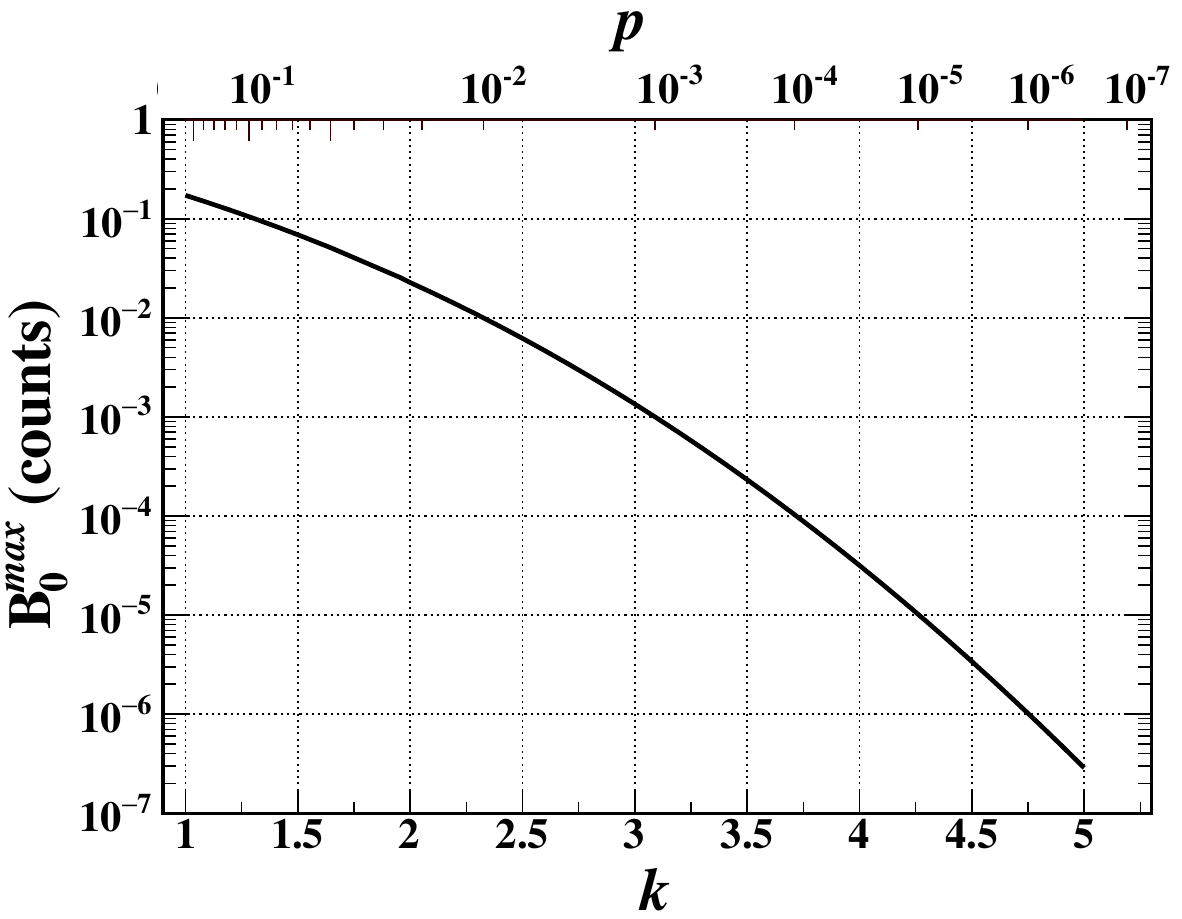}
\caption{
Variations of $\maxB0$ 
which satisfies the zero-background condition 
as a function of $k$ and $p$.
The contour is independent of $g$.
}
\label{fig::B0max}
\end{figure}


Poisson statistics is necessary in the
complete formulation of the problem.
For a given 
{positive} $\B0$ as input and
using $\P503s$ as illustration,
the Poisson distribution $\Poi (i;\mu)$
is constructed with mean $\mu {=} \B0$. 
{
Let $N_{obs}^{3 \sigma}$ be the minimal integer
number of observed events which provides
${\geq} 3 \sigma$ significance over a
predicted average background $\B0$. $N_{obs}^{3 \sigma}$
satisfies the following equation: 
\begin{equation}
\sum_{i=0}^{N_{obs}^{3 \sigma} - 1} \Poi ( i ; \B0 ) \geq ( 1 - p ) ~ .
\label{eq::B0}
\end{equation}
from which the value of $N_{obs}^{3 \sigma}$
can be determined.} The output $\S0$ is the
minimal signal strength
where a Poisson distribution with
$\mu {=} ( \B0 {+} \S0 )$
would give $N_{obs}^{3 \sigma}$ or more events with
$g {=}50\%$ probability:
\begin{equation}
\sum_{i=N_{obs}^{3 \sigma}}^{\infty} 
\Poi ( i ; { \B0 {+} \S0}  ) =  0.5 ~~.
\label{eq::B0S0}
\end{equation}

The required $\S0$ for criteria $\Pgk$
due to different $k$ and $g$
are shown in Figure~\ref{fig::DP-S0vsB0}.
The characteristic step-wise features are consequences of
the discrete nature in Poisson statistics $-$ 
only integer $\nobs$ are observed 
in one measurement. 
{
The steps for $\P503s$ and $\Prob903s$ occur at
the same $\B0$. This corresponds to the same required 
$N_{obs}^{3 \sigma}$ to meet
the ${\geq} 3 \sigma$ ($p {\leq} 0.00135$) criteria.
More $\S0$ events are necessary to establish a
positive signal in $\Prob903s$ than $\P503s$ 
when $g$ increases from  50\% to 90\% in Eq.~\ref{eq::B0S0}. 

Signal and background events are indistinguishable experimentally. 
The $\Pgk$ criteria and discreteness of Poisson statistics apply 
to $(\B0 {+} \S0)$. However, the useful information to experiments
is on the variation of $\S0$ with $\B0$. 
This explains the origin of the negative slopes between the steps
in Figure~\ref{fig::DP-S0vsB0}.
}

A particular case of interest is the 
``zero-background condition'' in which $\nobs {=} 1$ event
would qualify to be taken as a positive signal.
{
The maximum $\B0$ (denoted as $\maxB0$) 
where such conditions apply correspond 
to the ``first steps'' in Figure~\ref{fig::DP-S0vsB0}.
The dependence of $\maxB0$ on
$k$ and $p$ is depicted in Figure~\ref{fig::B0max}. 
The values of $\maxB0$ and $\S0$ under zero-background condition 
at different $\Pgk$ are summarized in Table~\ref{tab::zerobkg},
which illustrates the effects of $k$ and $g$. 

The values of $\maxB0$ $-$ and in general
the required $\nobs$ to establish positive signals at
${+} k \sigma$ excesses over background $-$ are described by
Eq.~\ref{eq::B0} and are therefore independent of the choice of $g$.
On the other hand, the required signal strength
$\S0$ at $\maxB0$ is given by
Eq.~\ref{eq::B0S0} and therefore has $g$-dependence.

}


\begin{table}
\caption{
Summary of the $\S0$ and $\maxB0$ values 
in counting-only analysis with complete Poisson statistics
at the zero-background condition
where $\nobs {=} 1$ event can establish a positive signal under
the criteria $\Pgk$.
}
\begin{center}
\begin{tabular}{|cc||cc|}
\hline
\multicolumn{2}{|c||}{ } & \multicolumn{2}{c|}{ Excess over background }\\
\multicolumn{2}{|c||}{ } & \multicolumn{2}{c|}{ $(~ k \sigma ~)$ }\\ \cline{3-4}
\multicolumn{2}{|c||}{ } & 
\multirow{2}{*}{~~~~~ ${+} 3 \sigma$ ~~~~~} & 
\multirow{2}{*}{~~~~~ ${+} 5 \sigma$ ~~~~~} \\ 
\multicolumn{2}{|c||}{ }  & & \\ \hline \hline
\multicolumn{2}{|c||}{ \multirow{2}{*}{ $ \maxB0 $ } }  &
\multirow{2}{*}{0.00135} & \multirow{2}{*}{$2.85 {\times} 10^{\mbox{-}7}$}  \\
\multicolumn{2}{|c||}{ } & &  \\ \hline
\multicolumn{2}{|c||}{ } & 
\multicolumn{2}{c|}{ \multirow{2}{*}{ \underline{ $\S0$ at $\B0 {<} \maxB0$ }}}\\ 
\multicolumn{2}{|c||}{ } & &  \\ 
& \multirow{2}{*}{ ~~ 50\% ~~ } & 
\multicolumn{2}{c|}{\multirow{2}{*}{0.69}} \\ 
Sample Fraction & & & \\ 
$ (~g~) $ & \multirow{2}{*}{ ~~ 90\% ~~ } & 
\multicolumn{2}{c|}{ \multirow{2}{*}{2.3} }  \\
 &  &  &  \\ \hline
\end{tabular}
\label{tab::zerobkg}
\end{center}
\end{table}


\subsection{Continuous Approximation to Poisson Distribution}
\label{sect::Poi_ContApp}


\begin{figure}
{\bf (a)}\\
\includegraphics[width=8.2cm]{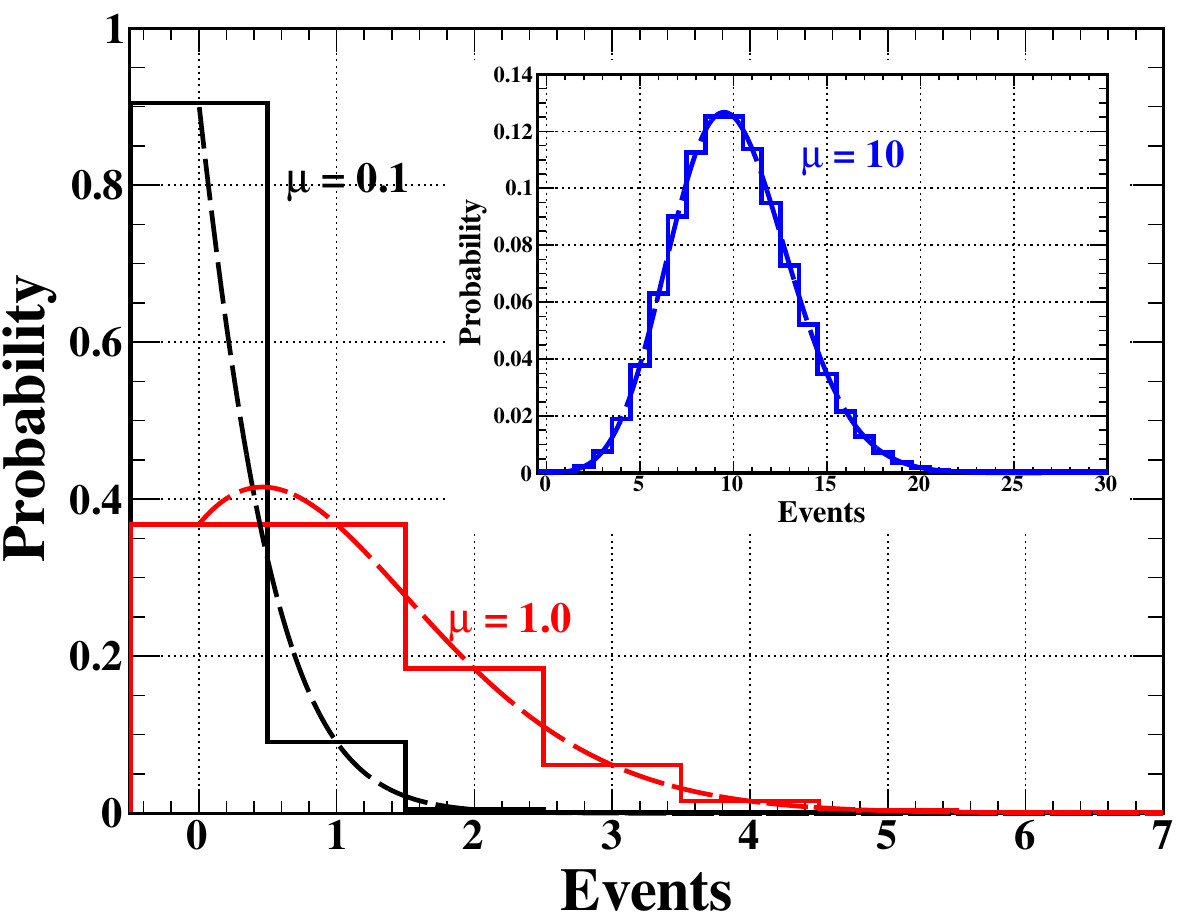}\\
{\bf (b)}\\
\includegraphics[width=8.2cm]{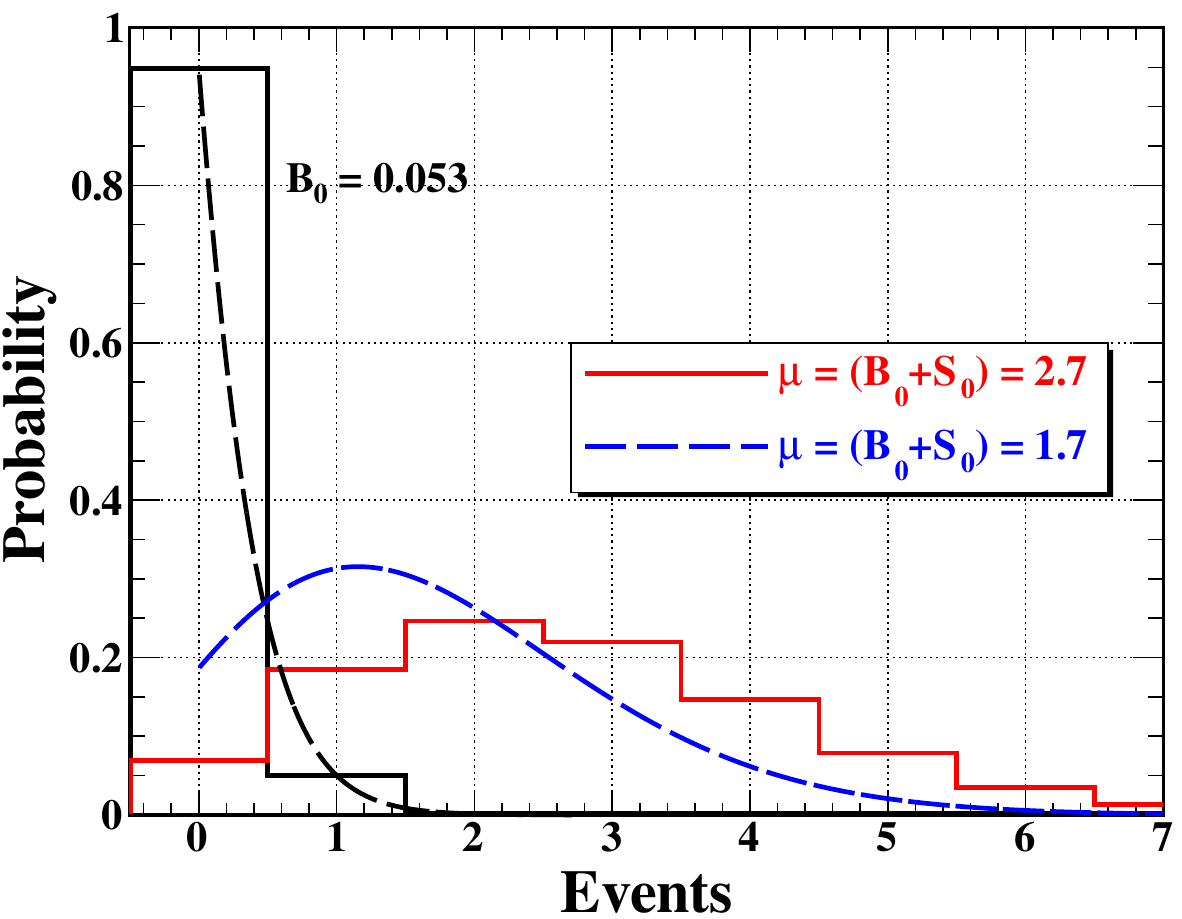}
\caption{
(a) Comparison of Poisson distribution in its complete formulation
and continuous approximations in the cases
of $\mu {=} \B0 {=} 0.1,1,10$.
(b)
The $\mu {=} ( \B0 + \S0 )$
distributions with $\P503s$ under both criteria
at $\B0 {=} 0.053$, showing their differences $-$
$\mu$=2.7(1.7) for complete Poisson (continuous approximation).
}
\label{fig::PvsC}
\end{figure}


Continuous approximations to the Poisson distributions
are derived by replacing  Eq.~\ref{eq::Poisson-Summation}
with the regularized incomplete gamma function:
\begin{equation}
\CPoi ( {\leq} C ; \mu ) = 
{  \frac{\Gamma ( C{+}1 ; \mu )}{\Gamma ( C{+}1 )}  ~~ ,}
\label{eq::Poisson-Continuous}
\end{equation}
where $C$ is generalized to be a
{continuous variable}.
The summations of Eqs.~\ref{eq::B0}\&\ref{eq::B0S0}
are replaced by Eq.~\ref{eq::Poisson-Continuous},
applicable for $\B0 {\geq} 0 $. 
This has been adopted to derive results to
the sensitivity projection problem~\cite{ijdeaor2016212,mbbprob:3,
Singh:2018lan,Continuous_Poisson:1,Continuous_Poisson:2,
Continuous_Poisson:3,RevModPhys.95.025002,PhysRevD.106.032012}.

The comparisons of the Poisson distribution
$\Poi ( n ; \mu {=} \B0 )$ and its continuous approximation
is depicted in Figure~\ref{fig::PvsC}a,
showing cases of 
{$\mu {=}  0.1,1,10 $ }
to illustrate behavior for different ranges.
For large $\mu$, the continuous formulation
approximates well to the discrete case, and
approaches the Gaussian distribution.


Only integer results are possible 
in counting measurements, so that the 
criterion ``${\geq} 3 \sigma$'' is mostly satisfied
as an inequality in the complete Poisson analysis. 
Illustrated in Figure~\ref{fig::PvsC}b
is an example of how $\S0$ would differ with the two formulations,
where the integration from zero of the histograms 
and dotted curves are different. 
The figure illustrates with the example of $\B0$=0.053. 
Individual experiments would require 
$\nobs {\geq} 3 (2)$  
to meet the ``${\geq} 3 \sigma$'' condition,
while $\P503s$ would imply average $\S0 {=} 2.64 (1.64)$ 
under complete Poisson counting and continuous approximation,
respectively.

\begin{figure} 
{\bf (a)}\\
\includegraphics[width=8.2cm]{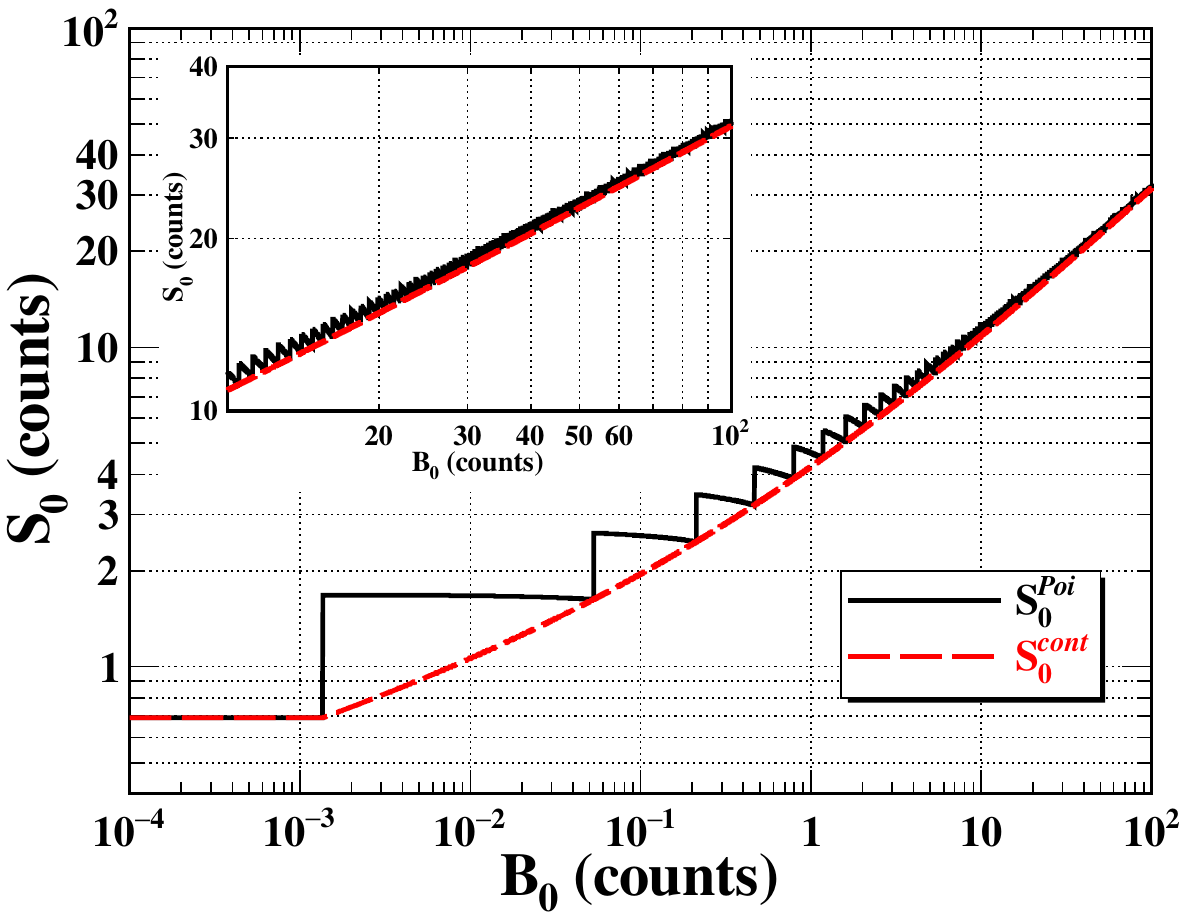}\\
{\bf (b)}\\
\includegraphics[width=8.2cm]{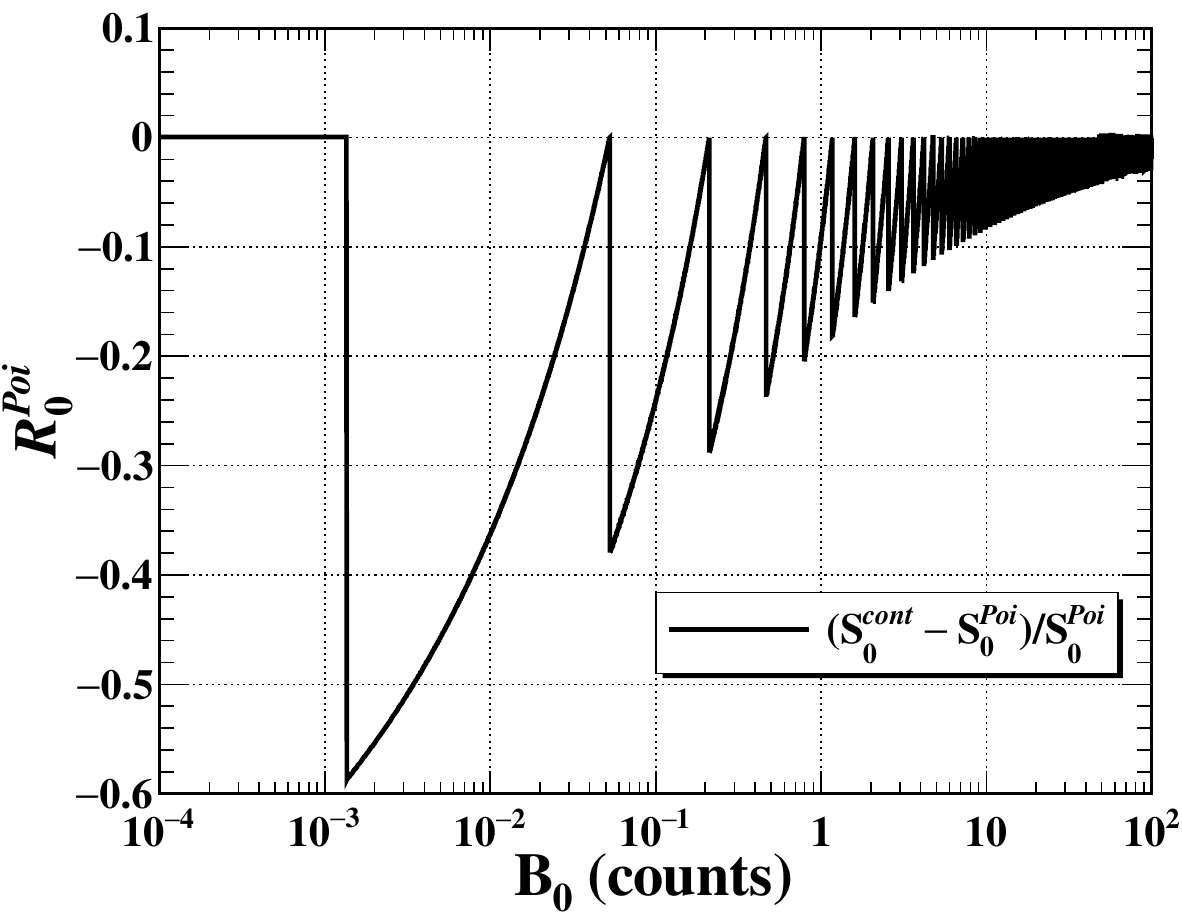}
\caption{
(a) Comparison between counting-only with complete Poisson
($\SPoi$) and continuous approximation ($\Scont$)
in defining the $\P503s$ sensitivity.
(b) Relative change ($R_{0}^{Poi}$) of $\Scont$
relative to $\SPoi$.
}
\label{fig::Discrete-B0}
\end{figure}

Results on the dependence of
$\S0$ versus $\B0$ from both formulations
are depicted in Figure~\ref{fig::Discrete-B0}a.
The $\S0$ derived with complete Poisson statistics ($\SPoi$)
is always larger than that from continuous approximation ($\Scont$),
except at where equality (${=} 3 \sigma$) is met.
The fractional decrease is depicted in 
Figure~\ref{fig::Discrete-B0}b by the black line,
where $R_{0}^{Poi} {=} ( {\Scont} {-} {\SPoi} ) / {\SPoi}$.
It can be seen that 
the continuous approximation always underestimate 
the necessary strength to establish a signal.
The deviation can be as much as 60\% 
at low background ($\B0 {\sim} 10^{\mbox{-}3}$),
but reduced to within 3\%
at large statistics of $\B0 {\gtrsim} 100$.


\section{Likelihood Analysis}
\label{sect::maxlikelihood}

In Section~\ref{sect::counting_analysis_poisson}, 
event count is used as ``test statistic''~\cite{loglikelihoodratio:3,Workman:2022ynf}.
This is a straight forward choice for experiments that 
measure a single integer value 
as the only output.
However, in experiments with measurements of multiple variables,
the Poisson counting method is insufficient to
extract complete information available in
the signal and background.
An alternative and more comprehensive formulation of
test statistic is therefore necessary.

A test statistic is a mapping from an experimental outcome with multiple
values to a single real number. 
The optimal test statistic is the likelihood ratio, 
following the Neyman–Pearson lemma~\cite{neyman-pearson-lemma}.

In this analysis, we adopt log likelihood ratio (LLR)
in Sections~\ref{sect::maxlikelihood-formulation}\&\ref{sect::extendedikelihood-formulation}
to be the test statistic where 
{$S {=} \S0$}
is a free parameter and 
{$B {=} \B0$} is fixed.
For cases where the uncertainties in $B$ are considered
as in Sections~\ref{sect::delta-bkg}\&\ref{sect::0nubb}, 
a variant of LLR with additional ``nuisance parameter''
(called log profile likelihood ratio) is used.

\subsection{Formulation and Single Integer Counting}
\label{sect::maxlikelihood-formulation}

The counting-only likelihood function is given by:
\begin{eqnarray}
\label{eq::likelihood-counting}
\mathscr{L}_C & ~ \equiv ~ & \mathscr{L}( S |N, B )  \\
& ~ = ~ & \frac{e^{-( B + S )}( B + S )^{N}}{N!}. \nonumber
\end{eqnarray}
Following conventional notations of
Refs.~\cite{loglikelihoodratio:1,Workman:2022ynf},
the LLR, denoted by $q_{0}$, is defined as
\begin{equation}
\q0  ~~ {\equiv} ~~ 
{t ( S {=} 0 ) }~ = ~
- 2 ~ { \rm ln }  \left[ \frac{\mathscr{L}(S{=}0)}{\mathscr{L}(\hat{S})} \right] ~ ,
\label{eq::q0}
\end{equation} 
in which $\hat{S}$ is the value of $S {\in} (0,\infty)$ that $\mathscr{L}(S)$
is maximized for given $N$ and at a fixed $B {=} \B0$ value. 
{
The $q_{0}$ is defined as a test statistic ($t$) which
serves as the foundation of a statistical test under the special
case where $S{=}0$.
}

We are interested in this work 
to quantitatively assess the significance of a measurement 
in supporting a discovery scenario.
Accordingly, the data set has to be tested against the
{\it null hypothesis} ($\Hnull$) case of  $S {=} 0$.
Consistent data set of $\Hnull$ with $S {=} 0$ will give
$\q0 {\rightarrow} 0$
whereas large $\q0$-values imply deviation from $\Hnull$.
The {\it alternative hypothesis} ($\H1$) 
characterizes the case with $S {=} \S0 {>} 0$,
where $\S0$ is the mean signal strength.
If a significant fraction of a data set generated by $\H1$
gives large $\q0$-values,
$\Hnull$ would have to be rejected.

The probability distributions of $\q0$ {for} given $\B0$
are evaluated from data sets simulated with
$\mathscr{L}_{C}$ having $N {=} \B0 {+} \S0$ events:
(i) $\nll-Pq0H0$ corresponding to $\Hnull$ with data at $\S0 {=} 0$, and
(ii) $\alt-Pq0H1$ corresponding to $\H1$ with data at
{non-zero} $S {=} \S0 {>} 0$.

Standard statistics variables are adopted
to quantify statistical consistency with hypotheses
in $\nll-Pq0H0$ and $\alt-Pq0H1$. 
Data with $\q0 {<} \talpha $ are considered to be 
within the ``acceptance interval'' consistent
with $\Hnull$, where $\talpha$ is a boundary to 
the ``size of test''\cite{loglikelihoodratio:3} 
(also called the Type-1 error and denoted as $\alpha$), 
a pre-defined value corresponding to 
the probability that the data set which is inconsistent with $\Hnull$,
or equivalently when $\q0$ is rejected to be $\Hnull$:
\begin{equation}
\alpha \equiv \int_{\talpha }^{\infty} ~ \nll-Pq0H0 ~ d \q0 ~ .
\label{eq::alpha_size_of_test}
\end{equation}
The ``power of test''~\cite{loglikelihoodratio:3} 
corresponds to ($1 {-} \beta$), where $\beta$ (also called the Type-2 error)
is the probability of $\q0$ within the acceptance region of $\Hnull$
in the scenario where the hypothesis $\H1$ is true.
It can be expressed as:
\begin{equation}
\beta \equiv \int_{0}^{\talpha } ~ \alt-Pq0H1 ~ d \q0 ~~.
\label{eq::beta_power_of_test}
\end{equation}


{
In counting experiments,
integrations in Eqs.~\ref{eq::alpha_size_of_test}\&\ref{eq::beta_power_of_test}
should be replaced by summations, such that: 
\begin{eqnarray}
\alpha & ~ \geq ~ & \sum_{\q0 \geq \talpha} \nll-Pq0H0 ~ , ~ {\rm and}   \nonumber \\
\beta & ~ = ~ &  \sum_{\q0 \leq \talpha} \alt-Pq0H1  ~ .
\label{eq::discrete_alpha_beta}
\end{eqnarray}
As a result of discreteness relevant and crucial to low-statistics counting,
$\alpha$ in general cannot be exactly equal to, and should instead over-cover,  
the  ``size of test''. 
Therefore, $\alpha$ should be defined instead as an inequality.
On the contrary, the $\beta$-condition depends on the 
mean signal strength $\S0$ which is a real number,
so that it can be satisfied as an equality.
}


The criteria $\Pgk$ defined in this work
corresponds to the matching of $p {=} \alpha$ and $g {=} ( 1 {-} \beta )$
to the standard statistical variables.
Accordingly, $\P503s$ implies the choice of $\talpha$
which leads to $p {=} 0.00135$ for $\nll-Pq0H0$ 
with $\q0 {\in} [ 0 , \talpha ]$.
{
Experiments with $\q0 {\in} [\talpha , \infty ]$
are inconsistent with $\Hnull$.
}
In addition, there is $( 1 {-} \beta ) {=} 50\%$ 
probability  to have $\q0 {\in} [ \talpha , \infty ]$
in $\alt-Pq0H1$  so that the experiment is
recognized to have observed positive signals.

\begin{figure}
{\bf (a)}\\
\includegraphics[width=8.0cm]{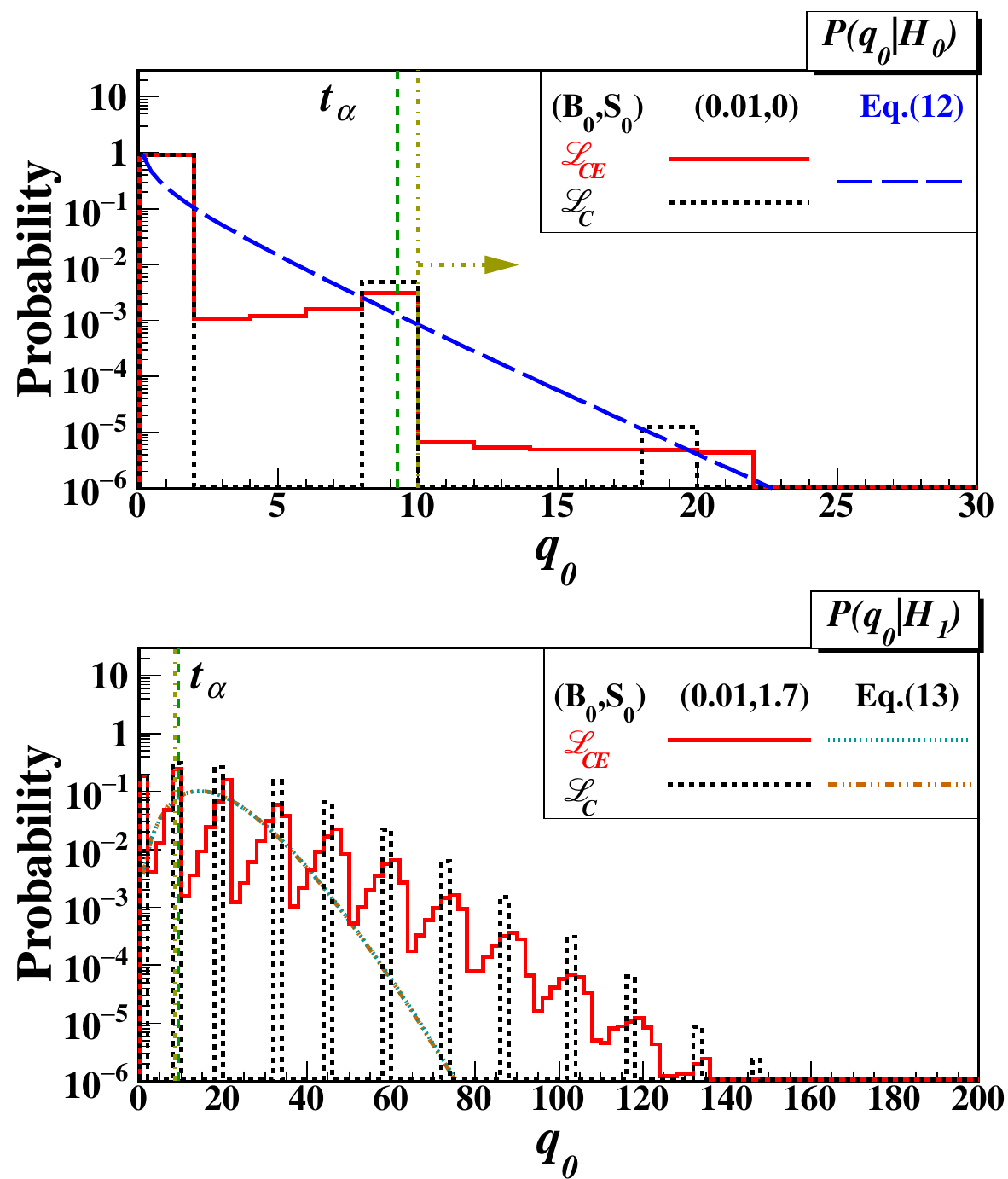}\\
{\bf (b)}\\
\includegraphics[width=8.2cm]{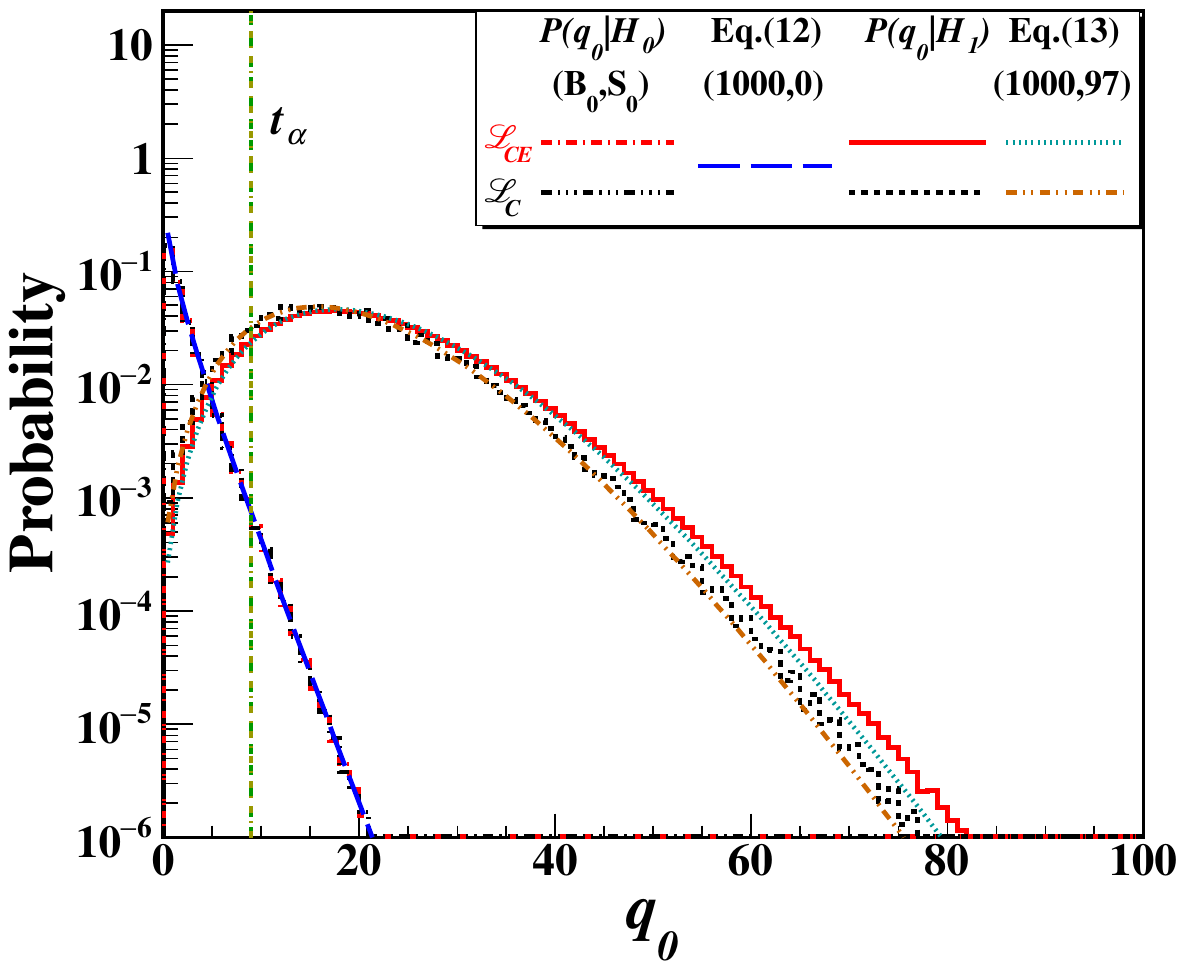}
\caption{
Distributions of test statistic $\q0$ for
simulated data with
null [$\nll-Pq0H0$] and alternative [$\alt-Pq0H1$] hypotheses
{for} negligible $\B0$ uncertainties:
(a) depicts a low-$\B0$ case with
$( \B0 , \S0 ) {=} ( 0.01 , 0 ) $ and $( 0.01 , 1.7 ) $
while (b) is a large-$\B0$ case with
$( \B0 , \S0 ) {=} ( 1000 , 0 ) $ and $( 1000 , 97 ) $.
The acceptance criteria specified by $\talpha $ are displayed.
The approximations of
Eqs.~\ref{eq::chi_square_distribution}\&\ref{eq::noncenter_chi_square_distribution}
are superimposed, verifying that
they match $\nll-Pq0H0$ and $\alt-Pq0H1$ {for}
large but fail {for} small $( \B0 , \S0 )$.
}
\label{fig::MLR-B0}
\end{figure}

As a result of the discreteness of single-value integer counting,  
the count to $\q0$ mapping is always one to one at $\hat{S} {>} 0$.
Examples of $\nll-Pq0H0$ and $\alt-Pq0H1$ distributions
for LLR counting analysis with $\mathscr{L}_{C}$
are shown in Figures~\ref{fig::MLR-B0}a\&b, 
which describe cases of low- and high-statistics, respectively.

In the absence of additional measurables,
the LLR analysis on $\mathscr{L}_C$ 
results in  $\S0 [ {\mathscr{L}_C} ]$ {(signal strength of
counting-only LLR analysis)} which are identical to $\SPoi$
derived by the complete Poisson counting analysis.
The counting-only results of Figure~\ref{fig::DP-S0vsB0}
and Figures~\ref{fig::Discrete-B0}a\&b can
be derived by both formulations in 
Section~\ref{sect::poisson-formulation}
and 
Section~\ref{sect::maxlikelihood-formulation}.


\subsection{Extended Likelihood with Additional Measurables}
\label{sect::extendedikelihood-formulation}


In realistic applications, 
such as $\0nubb$ experiments
to be discussed in Section~\ref{sect::0nubb}, 
the observables
typically include energy. 
Without loss of generality, 
we take energy of an event
to be the additional available observable.
The studied scenario
is with signal events having
known mono-energetic $\E0$
smeared by experimental resolution
characterized by Gaussian peaks
with known width $-$ 
RMS and FWHM (full-width-half-maximum) 
denoted by $\sigmaE0$ and 
$\DeltaE0 {(\equiv 2.355{\times}\sigmaE0)}$, respectively.
The background is known and is a constant independent with energy,
characterized by $\B0$ and $\dB$ denoting, respectively,
the expected background count and its RMS uncertainty.
An RoI has to be specified in the analysis in
such experiments, in which additional energy measurements
are available. A natural choice would be 
$( \E0 {\pm} \Nsigma \sigmaE0 )$ where the variable $\Nsigma$ would
parametrize the interval width of the RoI.
Background is then quantified as $( \B0 / \sigmaE0 )$ 
in units of counts-per-RMS, 
as compared to the exclusive counting-only cases of $\B0$(counts) 
in Section~\ref{sect::counting_analysis_poisson}.

In the limit of $\dB {\ll} B$ 
where the background is accurately predicted, 
the likelihood function of
a signal $S$ given a known background profile $B$
and a data set $\mathbb{E}$
with $N$ events with measured energy $E_i (i=1,N)$
can be described by the
extended likelihood function:
\begin{eqnarray}
\label{eq::extended_likelihood}
\mathscr{L}_{CE} & ~ \equiv ~ & \mathscr{L}( S | \mathbb{E} , B )  \nonumber \\
& ~ =  ~ & \frac{e^{-( B + S )}( B + S )^{N}}{N!}  \times  \\ 
 & & \prod_{i=1}^{N}
\left[ \frac{ B  \cdot {f_{B}(E_{i})}+ S  \cdot f_{S}(E_{i})}{( B + S )} \right] ~ , \nonumber
\end{eqnarray}
where $f_{B}$ and $f_{S}$ are normalized probability density functions
of, respectively, background and signal,
such that
$\int_{\rm RoI}{f_{B} (E) dE} {=} 1$ and 
$\int_{\rm RoI}{f_{S}(E)dE} {=} 1$.

In our adopted $\0nubb$-inspired scenario, 
$B {=} \B0$ and $f_B$ is a constant independent of energy,
while $f_S$ is a Gaussian with known mean and width.
Results on $\mathscr{L}_{CE} ( S )$ from Eq.~\ref{eq::extended_likelihood}
is independent on the choice of RoI, so long as it covers 
the entire signal region$-$
$\RoI ( \mathscr{L}_{CE} ) {=} \E0 {\pm} 4 \sigmaE0$ is selected in this analysis,
with which $\effRoI ( \mathscr{L}_{CE}) {=} 0.9999$.

The LLR of Eq.~\ref{eq::q0}
is selected~\cite{feldman-cousins} as the 
test statistic ($\q0$)~\cite{loglikelihoodratio:1,loglikelihoodratio:2,loglikelihoodratio:3,Workman:2022ynf}.
Unlike those from counting analysis of Eq.~\ref{eq::q0},
probability distributions of $\q0$ do not have analytical form
for both the $\Hnull$ and $\H1$ hypotheses,
and have to be generated by simulation.
Approximation methods can be used in the
special cases of large samples, as discussed in 
Section~\ref{sect::q0-approximation}.

The case of $\dB {\ll} B$ was first studied.
A total of 50-million experiments are generated for each $\mathbb{E}$ 
with different input values of $\S0$.
The number of background ($N_B$) and signal ($N_S$) events
for individual experiment follow Poisson statistics:
$\Poi(N_{B}|\B0 )$ and $\Poi(N_{S}|\S0 )$, respectively,
while their energy distributions follow {$f_B (E)$} and $f_S (E)$ within the RoI.
{The total number of events}, $N {=} N_B {+} N_S$, varies with each experiment.
The $\hat{S}$-values which maximize $\mathscr{L}$ 
for individual experiments are derived, from which
the $\q0$-values of Eq.~\ref{eq::q0} are evaluated.
Their distributions over large number of experiments
in $\nll-Pq0H0$ and $\alt-Pq0H1$ 
corresponds to
the probability densities where $\q0$ 
is consistent with $\Hnull$ and $\H1$, respectively.


Displayed in Figure~\ref{fig::MLR-B0}a are
distributions of $\nll-Pq0H0$ and $\alt-Pq0H1$ as 
functions of $\q0$ in both 
$\mathscr{L}_{C}$ and $\mathscr{L}_{CE}$
{for} a low-statistics case, 
where $( \B0 , \S0 ) {=} ( 0.01 , 0 ) $ and $( 0.01 , 1.7 ) $.
The analogous high-statistics case at
$( \B0 , \S0 ) {=} ( 1000 , 0 ) $ and $( 1000 , 97 ) $
is shown in Figure~\ref{fig::MLR-B0}b.
As additional energy information is incorporated to
the analysis,
$\nll-Pq0H0$ and $\alt-Pq0H1$ are 
{\it smeared out} in low statistics,
while changes are minor in high statistics.


The $\talpha$-values corresponding 
to ${\geq} 3 \sigma$ upward excesses from $\Hnull$
are marked in Figures~\ref{fig::MLR-B0}a\&b. 
In particular in the 
high statistics limit where $\B0 {=} 1000$ in
Figure~\ref{fig::MLR-B0}b, 
$\nll-Pq0H0$ approximates to 
$\chi^2$-distribution and $\talpha {\rightarrow} 9$. 


\subsection{Approximate Distribution of $\q0$ for Large-Samples}
\label{sect::q0-approximation}

Following the formulation by Wilks~\cite{wilks} 
and Wald~\cite{wald}, 
$\nll-Pq0H0$ or $\alt-Pq0H1$ can be simplified
in the large-sample limit,
where Poisson distributions can be {approximated by} Gaussian.
Computing resources in simulations can therefore be saved
by the use of analytic equations when results are evaluated 
from input spanning large parameter space.

{When $S \geq 0$}, $\nll-Pq0H0$ is given by
half $\chi^2$-distribution 
for one degree of freedom 
plus a half $\delta$-function:
\begin{equation}
\label{eq::chi_square_distribution}
\nll-Pq0H0 \approx \frac{1}{2}\delta(\q0)
+ \frac{1}{2}\frac{1}{\sqrt{2\pi}}\frac{1}{\sqrt{\q0}}e^{-\q0/2} ~ ,
\end{equation}
while $\alt-Pq0H1$ is described 
by non-central $\chi^2$-distribution for one degree of freedom:
\begin{eqnarray}
\label{eq::noncenter_chi_square_distribution}
\alt-Pq0H1 & \approx & (1-\Phi(\sqrt{\Lambda}))\delta(\q0)  \\
& & + \frac{1}{2}\frac{1}{\sqrt{2\pi}}
\frac{1}{\sqrt{\q0}}e^{-(\sqrt{\q0}-\sqrt{\Lambda})^{2}/2} ~ , \nonumber
\end{eqnarray}
where $\Lambda$ is the non-centrality parameter, 
and $\Phi$ is cumulative Gaussian distribution.
The $\Lambda$ is the $\q0$ value of 
most probable $-$ that is, 
Asimov $-$ data set~\cite{loglikelihoodratio:1}.


Binned likelihood function is used
in the evaluation of $\Lambda$:
\begin{equation}
 \label{eq::binned_expected_value}
\mathscr{L}( S | \{n_{i}\} , B )  \approx 
\prod_{i=1}^{n} \Poi(n_{i}|F(E_{i}|S,B)) ~ , 
\end{equation}
where
\begin{equation}
F(E_{i}|S,B)  =  \left[ ~ B  \cdot {f_{B}(E_{i})} + 
S  \cdot f_{S}(E_{i}) ~ \right] \cdot w(E_i)  
\end{equation}
{
is the expected counts in the $i^{th}$-bin 
with bin size $w ( E_i )$,
$n_{i}$ is the measured count and 
$E_{i}$ is the mean energy. 
}
We note that likelihood expression of Eq.~\ref{eq::binned_expected_value} 
differs from Eq.~\ref{eq::extended_likelihood}
by a scaling constant which is canceled out 
when taking likelihood ratio.

The Asimov data set is therefore 
the expected count in each bin:
\begin{equation}
n_{i} = F(E_{i}|\S0,\B0) ~ ,
\label{eq::define_asimov}
\end{equation}
where $\S0$, $\B0$ are the input values
to generate the simulated data.
Accordingly, the $\Lambda$-value is the likelihood ratio:
\begin{equation}
\Lambda \approx - 2 ~ {\rm ln} \left[ \frac{\mathscr{L}( S{=}0 | B, n_{i} {=} 
F(E_{i}|\S0,\B0) )}{\mathscr{L}( \hat{S} | B, n_{i} {=} F(E_{i}|\S0,\B0) )} \right] ~ ,
\label{eq::define_lambda}
\end{equation}
with the $n_{i}!$ factorial terms 
in denominator and numerator canceled out.

The approximations of
$\nll-Pq0H0$ and $\alt-Pq0H1$ by
Eqs.~\ref{eq::chi_square_distribution}\&\ref{eq::noncenter_chi_square_distribution}
in the low- and high-statistics regimes
are superimposed in Figures~\ref{fig::MLR-B0}a\&b, respectively.
It can be seen that {for} the high-statistics limit, 
the approximations match well with the simulation results of $\nll-Pq0H0$ and $\alt-Pq0H1$,
but they deviate significantly {in} the low-statistics regimes. 


\begin{figure} 
{\bf (a)}\\
\includegraphics[width=8.2cm]{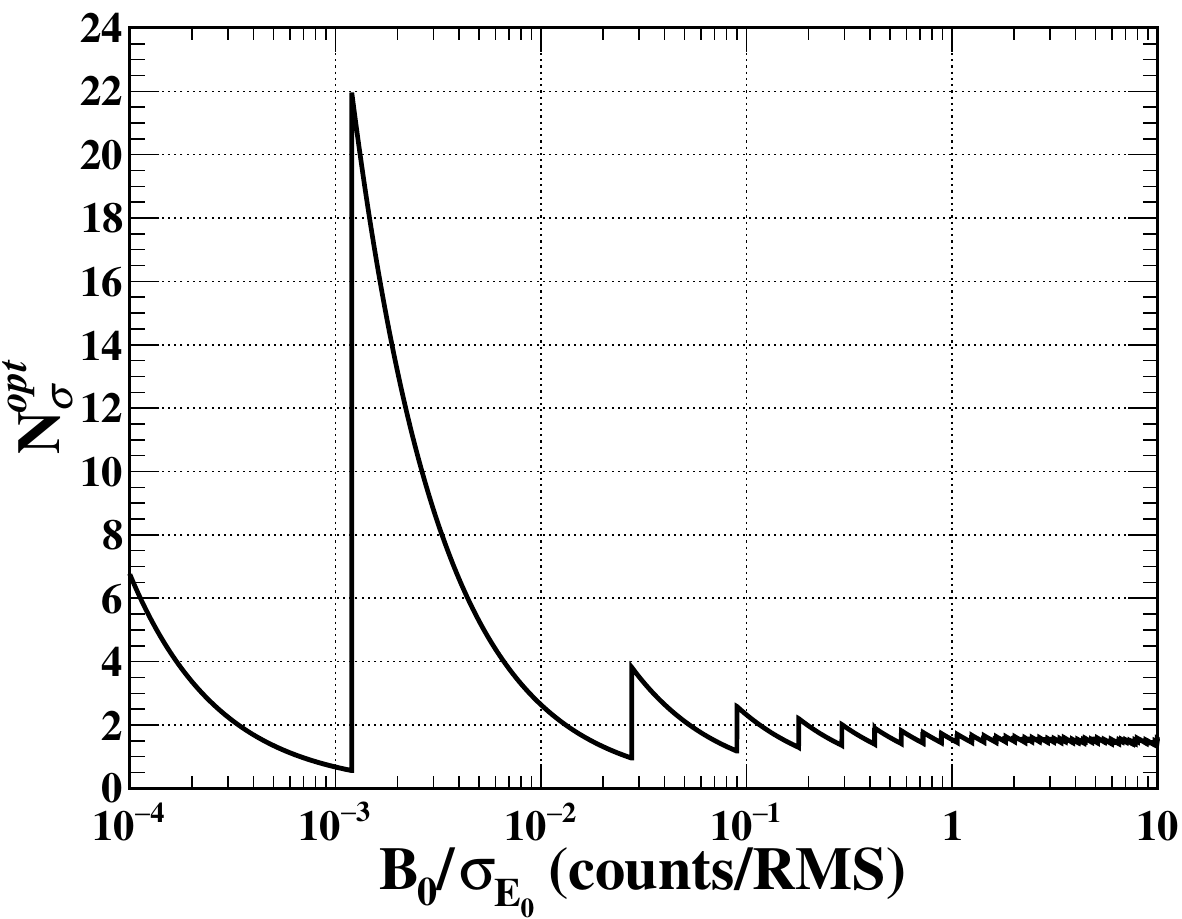}\\
{\bf (b)}\\
\includegraphics[width=8.2cm]{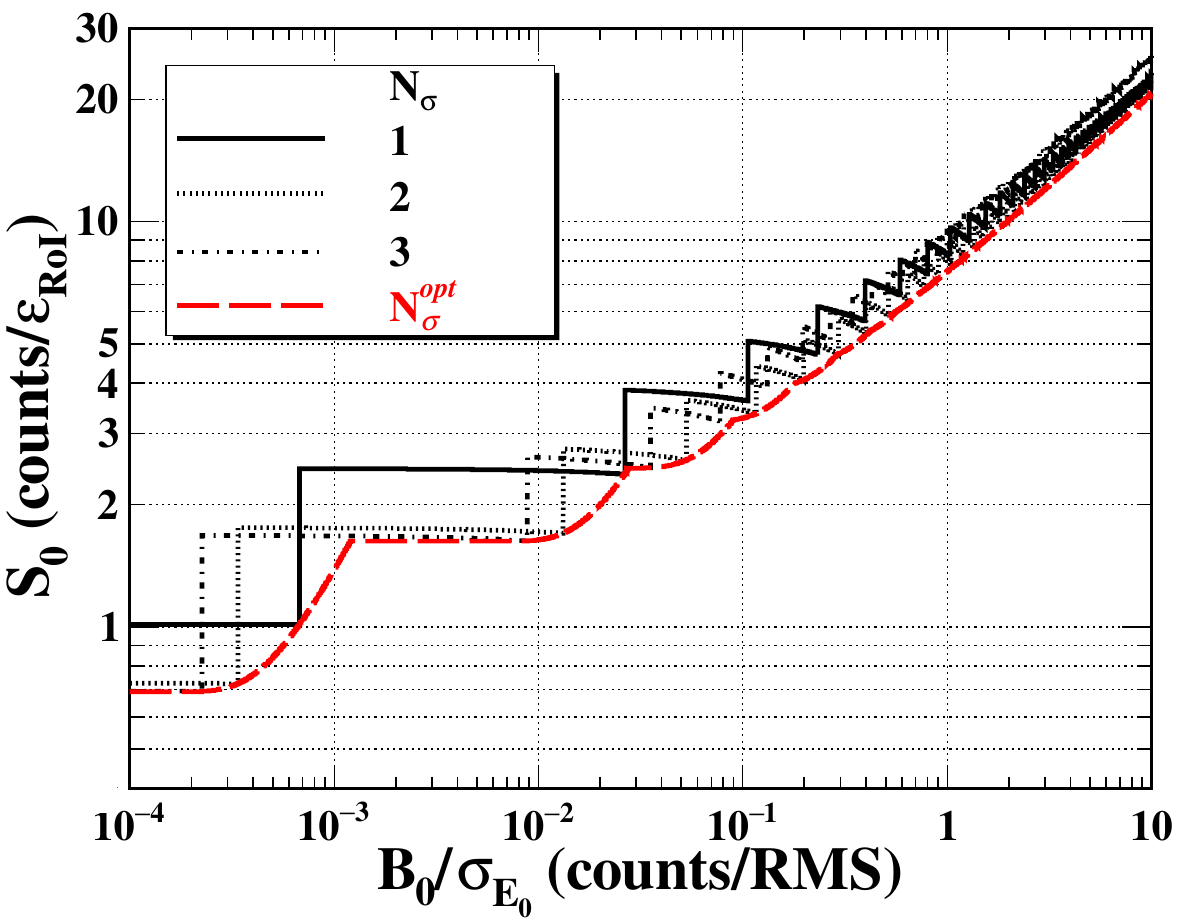}
\caption{
(a) Variation of  $\Noptsigma$
versus $( \B0 / \sigmaE0 )$ in counting-only analysis
{for} which the required $\S0$
to satisfy $\P503s$ are at minimum.
The RoIs are defined by intervals $\E0 {\pm} \Nsigma \sigmaE0$.
(b) Comparison of $\S0$ versus $( \B0 / \sigmaE0 )$
at $\Noptsigma$ with
those at fixed $\Nsigma {=} 1,2,3$.
}
\label{fig::optimal}
\end{figure}


\begin{figure} 
{\bf (a)}\\
\includegraphics[width=8.2cm]{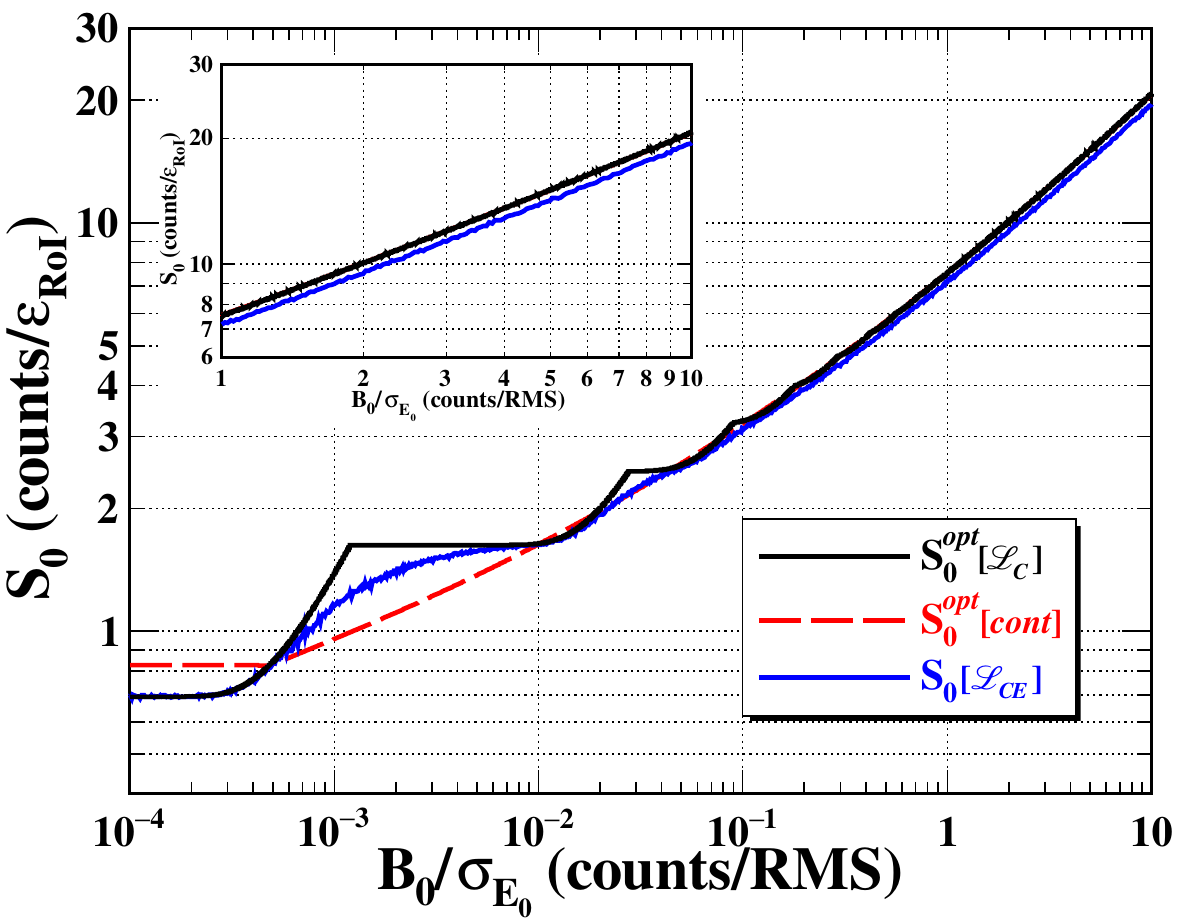}\\
{\bf (b)}\\
\includegraphics[width=8.2cm]{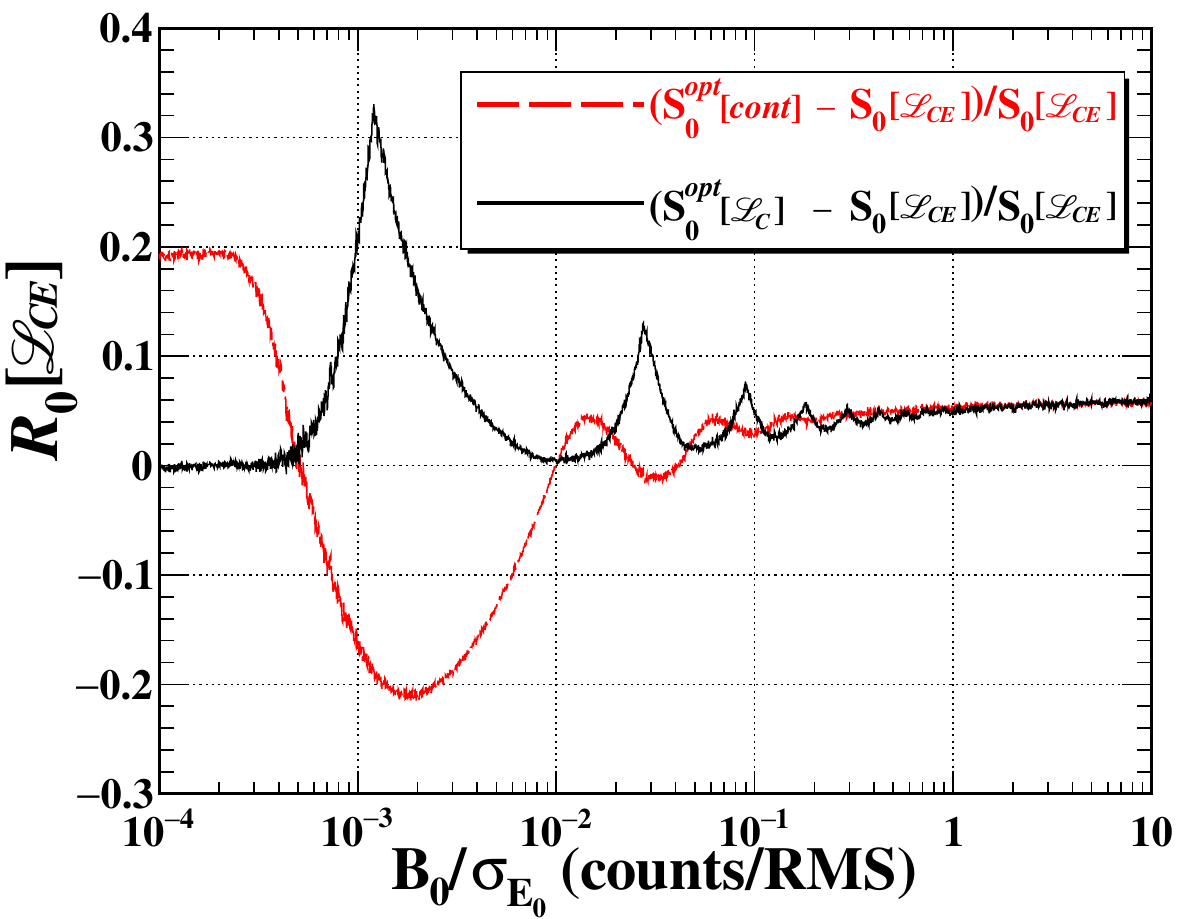}
\caption{
Sensitivities of $( \S0 / \effRoI )$
as a function of $( \B0 / \sigmaE0 )$:
(a)
On $\mathscr{L}_{CE}$ by LLR analysis
with complete information incorporated,
choosing
$\RoI ( \mathscr{L}_{CE} ) {=} \E0 {\pm} 4 \sigmaE0$.
These are compared with those
counting-only analysis via $\mathscr{L}_{C}$
and continuously approximation
{for} the optimal RoI of $\E0 {\pm} \Noptsigma \sigmaE0$.
(b) The deviations of
 $\Sopt [ {\mathscr{L}_{C}} ]$ and $ \Sopt [ cont ]$
relative to
$\S0  [{\mathscr{L}_{CE}} ]$, denoted as $R_{0} [ {\mathscr{L}_{CE}} ]$.
}
\label{fig::comparison}
\end{figure}


\subsection{Comparison between Counting and Extended Likelihood Analysis}
\label{sect::comparison}

{

Taking experiments where where both counts and energy are measured,
the required $\S0$-strength to achieve the $\P503s$ discovery potential criteria 
are derived. 
Several analysis schemes are compared:
(i) with the LLR analysis using $\mathscr{L}_{CE}$ of Section~\ref{sect::extendedikelihood-formulation}
exploiting both information, denoted $\S0 [ \mathscr{L}_{CE} ]$,
(ii) with a counting-only analysis via $\mathscr{L}_{C}$
of Section~\ref{sect::maxlikelihood-formulation}
discarding the available energy information, 
denoted $\S0 [ \mathscr{L}_{C} ]$
(this is equivalent to $\SPoi$ of Section~\ref{sect::poisson-formulation}
when the RoI intervals and $\effRoI$ are taken into account~\cite{TEXONO:PRD:2020}), and
(iii) with a counting-only analysis the continuous approximation
of Section~\ref{sect::Poi_ContApp}~\cite{ijdeaor2016212,mbbprob:3,Singh:2018lan,Continuous_Poisson:1,
Continuous_Poisson:2,Continuous_Poisson:3,RevModPhys.95.025002,PhysRevD.106.032012},
denoted $\S0 [ cont ]$.

As noted in Section~\ref{sect::extendedikelihood-formulation},
the sensitivities on $\S0 [ \mathscr{L}_{CE} ] $ 
is independent on the choice of RoI, so long as
$\effRoI {\simeq} 1$,  such as 
$\RoI ( \mathscr{L}_{CE} ) {=} \E0 {\pm} 4 \sigmaE0$.
On the contrary, 
the counting-only analysis of (ii) and (iii) 
depend on the choice of RoI as parametrized by $\Nsigma$.  
The optimal $\Nsigma$ (denoted $\Noptsigma$) 
which gives minimal $\S0 [ \mathscr{L}_{C} ] ({\equiv}  \Sopt [ \mathscr{L}_{C} ] $)
and $\S0 [ cont ] ({\equiv}  \Sopt [ cont ] $)
can be evaluated.

The variation of $\Noptsigma$ as a function of  $( \B0 / \sigmaE0 )$
is displayed in Figure~\ref{fig::optimal}a.
As noted in Ref.~\cite{mbbprob:3} and verified in our results,
the choice of $\Noptsigma {=} 1.4$
is optimal at large $( \B0 / \sigmaE0 ) {\gtrsim} 1$.
The ranges of optimal RoIs {for} low $ ( \B0 / \sigmaE0 ) $
vary broadly due to large fluctuations {in} low counts
and the discreteness of Poisson statistics.
Depicted in Figure~\ref{fig::optimal}b is
$\Sopt [ { \mathscr{L}_{C} } ]$ superimposed with
the cases of fixed RoI {for} intervals $\E0 {\pm} \Nsigma \sigmaE0$
(where $ \Nsigma {=} 1,2,3 $) 
corresponding to 
$\effRoI {=} 68.3\%,95.5\%,99.7\% $, respectively.


The results of the three analysis schemes 
are compared in Figure~\ref{fig::comparison}a.  
The deviations of $\Sopt [ cont ]$
and $\Sopt [ \mathscr{L}_{C} ]$
relative to $\S0 [ \mathscr{L}_{CE} ]$  
are depicted in Figure~\ref{fig::comparison}b.

}

While the features can be expected,
the results verify and quantify that 
in experiments incorporating additional energy information,
the discovery potentials are enhanced 
due to ${\rm S}_0 [ {\mathscr{L}_{CE}} ] {\leq} \Sopt [ {\mathscr{L}_{C}} ]$
which implies  less events are required to establish positive signals.

At the low-statistics regime $[ ( \B0 / \sigmaE0 ) {\lesssim} 0.01]$,
this originates from that the $\P503s$ criteria can be satisfied
for all $\B0$ in $\mathscr{L}_{CE}$, which is not the case for
counting-only analysis in $\mathscr{L}_{C}$
due to ``over-coverage'' (the $p {=} 0.00135$ criteria cannot be met). 
At high statistics $[ ( \B0 / \sigmaE0 ) {\gtrsim} 0.1]$,
requirements for the energy values to match
a pre-defined Gaussian peak provide the dominant constraints.

At low $( \B0 / \sigmaE0 ) {\sim} 10^{\mbox{-}3}$,
the $\Sopt [ cont ]$ can underestimate 
the required strength of $\S0 [ \mathscr{L}_{CE} ]$  
by as much as 20\%.
The $\Sopt [ \mathscr{L}_{C} ] $,
on the other hand, can be overestimated by as much as 30\%
and is larger than $\S0 [ \mathscr{L}_{CE} ]$
for all $( \B0 / \sigmaE0 ) {>} 5 {\times} 10^{\mbox{-}4}$.
At large $( \B0 / \sigmaE0 ) {>} 1$,
both derivations with counting-only analysis give 
consistent results which overestimate 
$\S0 [ \mathscr{L}_{CE} ]$ by $\sim$6\%.


\subsection{Effects of Background Uncertainties}
\label{sect::delta-bkg}


\begin{figure} 
{\bf (a)}\\
\includegraphics[width=8.2cm]{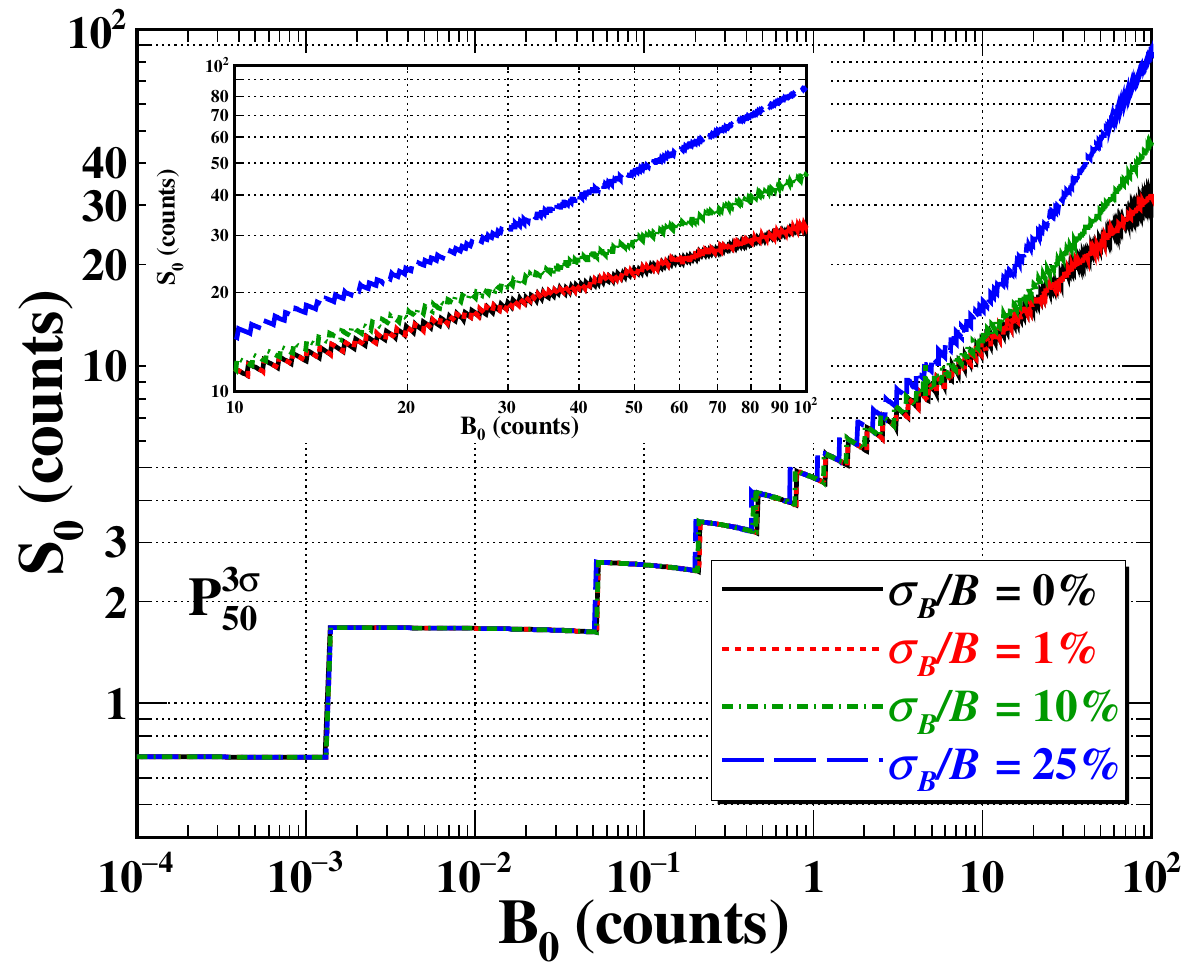}\\
{\bf (b)}\\
\includegraphics[width=8.2cm]{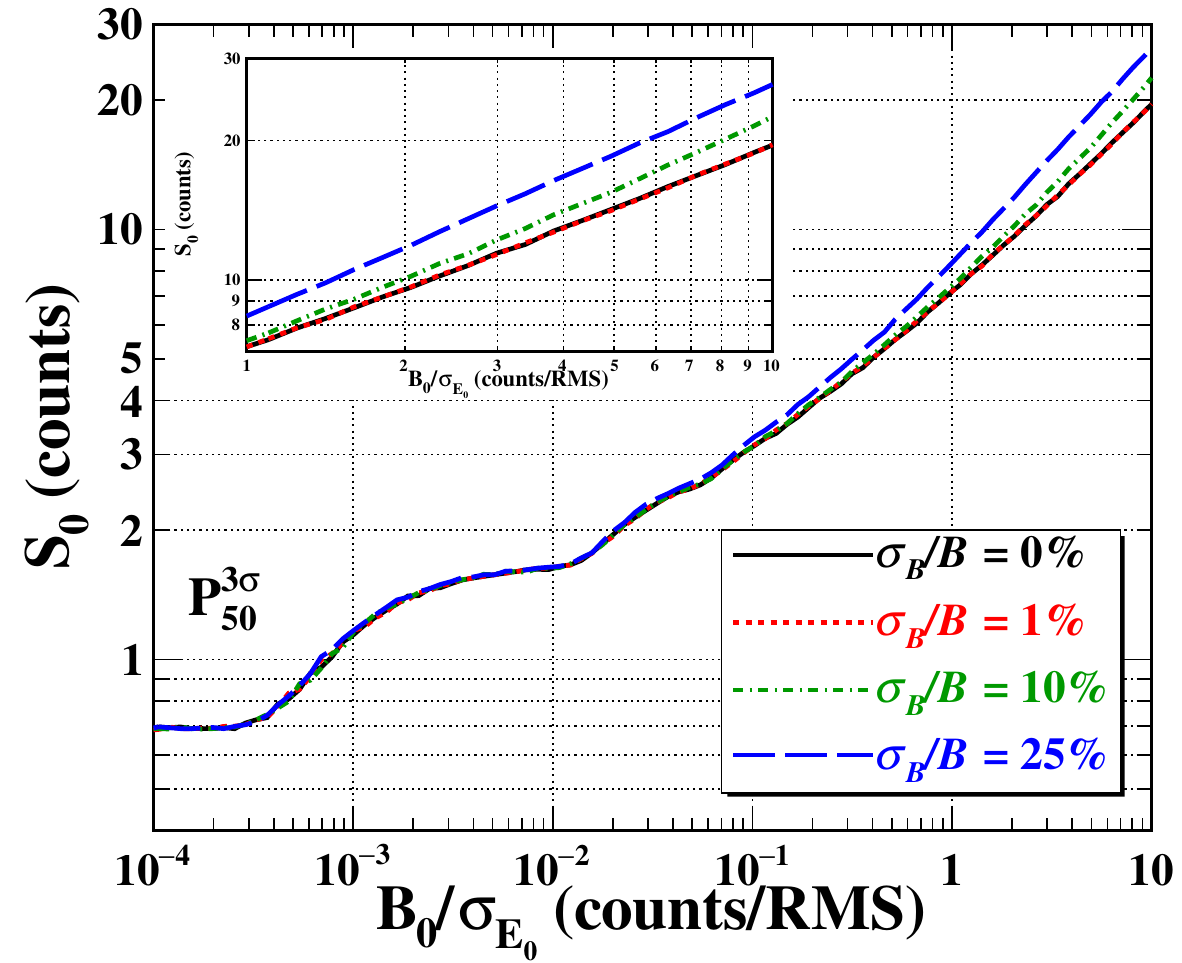}\\
{\bf (c)}\\
\includegraphics[width=8.2cm]{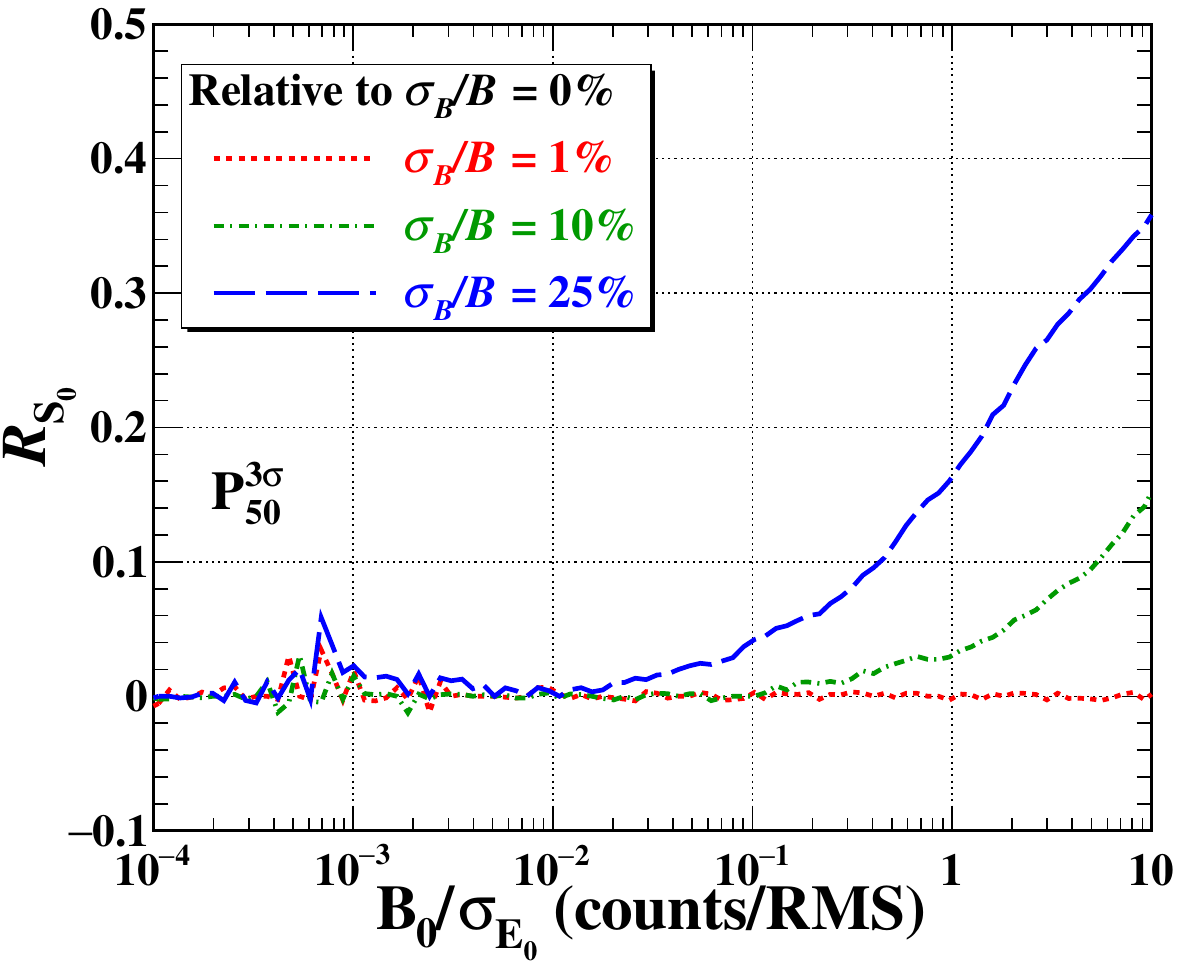}
\caption{
The effects on the sensitivities on $\S0$ defined by $\P503s$
due to background uncertainties ($\dB / B$)
(a)
in counting-only analysis with $\mathscr{L}_{C}$, and
(b)
in LLR analysis with energy information $( \mathscr{L}_{CE} )$
when the signal is an energy peak with Gaussian distribution,
and the selected $\RoI ( \mathscr{L}_{CE}) {=} \E0 {\pm} 4 \sigmaE0$.
{
(c)
The fractional increase of $\S0$ in $\mathscr{L}_{CE}$ (denoted as $R_{ \S0 }$) 
due to non-zero $ ( \dB / B )$
relative to the case of zero uncertainty.
}
}
\label{fig::deltaB}
\end{figure}


In realistic experiments,
the background $B$ is usually not precisely known
and can be characterized with an uncertainty $\dB$.
That background knowledge can be described as
auxiliary measurement channels 
(for instance, from simulations, prototype measurements,
extrapolations from non-RoI regions)
in the likelihood analysis.

The likelihood with an additional auxiliary channel 
can be described by another Poisson distribution $\Poi(n_{0}|{\tau}B)$,
and expressed as~\cite{loglikelihoodratio:1,loglikelihoodratio:2,loglikelihoodratio:3}
\begin{eqnarray}
\label{eq::extended_likelihood_with_errB}
\mathscr{L}_{CEB} & ~ \equiv ~ & \mathscr{L}( S,B | \mathbb{E} )  \nonumber \\
& =  & \frac{e^{-( B + S )}( B + S )^{N}}{N!}
\frac{e^{-{\tau} B  }( {\tau}B )^{n_{0}}}{n_{0}!}  \\
& & \times \prod_{i=1}^{N}
\left[ \frac{ B  \cdot {f_{B}(E_{i})}+ S  \cdot f_{S}(E_{i})}{( B + S )} \right] ~ , \nonumber
\end{eqnarray}
where $\tau$ is the ratio of data size of auxiliary measurement channel
relative to the main measurement channel, such that 
the RMS uncertainty in $B$ is
$\dB {=} \sqrt{{\tau}B}/\tau$.


{For non-zero} $\dB$,
additional values of $n_{0}$ for this auxiliary measurement
are generated {alongside}
$\Poi(N_{B}|\B0)$, $\Poi(N_{S}|\S0)$ as well as
data sets $\mathbb{E} ( \Hnull )$ and $\mathbb{E} ( \H1 )$
for Eq.~\ref{eq::extended_likelihood_with_errB}.
The LLR for test statistic
of Eq.~\ref{eq::q0}
is extended to:
\begin{equation}
\q0 \equiv t(S=0) {=} {\rm - 2 ~ ln } 
\left[ \frac{\mathscr{L}_{CEB} (S=0 ,\hat{\hat{B}})}{\mathscr{L}_{CEB} (\hat{S},\hat{B})} \right],
\label{eq::negative_log_likelihood_tauB}
\end{equation}
in which $\hat{\hat{B}}$ is, for given $\mathbb{E}$, 
the value of $B$ that 
maximizes $\mathscr{L}_{CEB} (S,B)$ in $B {\in} (0,\infty)$ 
at $S{=}0$ 
and $(\hat{S},\hat{B})$ is the $(S,B)$ that 
maximizes $\mathscr{L}_{CEB} (S,B)$ in $S {\in} (0,\infty)$ and $B {\in} (0,\infty)$.

The Asimov data set includes 
$n_{0} {=} {\tau}\B0$ in addition to the conditions
of  Eq.~\ref{eq::define_asimov}.
The binned likelihood function can be expressed as:
\begin{equation}
  \begin{split}
\mathscr{L}( S | \{n_{i}\} , B ) \approx  \hspace*{5cm} \\
\left[ ~ \prod_{i=1}^{n} \Poi(n_{i}|F(E_{i}|S,B)) ~ \right] \cdot
\Poi(n_{0}|{\tau}B)  ~~ .
  \end{split}
\label{eq::binned_likelihood_with_errB}
\end{equation}

An LLR analysis is performed on likelihood functions of 
$\mathscr{L}_{C}$ in Eq.~\ref{eq::likelihood-counting}
and
$\mathscr{L}_{CE}$ in Eq.~\ref{eq::extended_likelihood} 
with uncertainty term incorporated in
$\mathscr{L}_{CEB}$ in Eq.~\ref{eq::extended_likelihood_with_errB}.
Effects of a {non-zero} $( \dB / B )$ are studied through
the $\q0$ distributions for $\nll-Pq0H0$ and $\alt-Pq0H1$ {in}
both low and high statistics, analogous to Figures~\ref{fig::MLR-B0}a\&b.
The expected signal counts that 
meet the $\P503s$ criteria {for} different $( \dB / B )$ values
to the count-only and count-plus-energy cases, respectively,
are depicted 
in Figures~\ref{fig::deltaB}a\&b.
{
The fractional increase of $\S0$ in $\mathscr{L}_{CE}$ 
due to non-zero $ ( \dB / B )$
relative to the case of zero uncertainty
is given in Figure~\ref{fig::deltaB}c.
}

It can be seen 
that at the low-statistics regime 
($\B0 {<} 1$ within RoI=$\E0 {\pm} 4 \sigmaE0$) 
the effects of $\dB$ are negligible. 
The {reason} is that statistical fluctuations
of small numbers in a single measurement
dominate over the inadequate knowledge of 
the background.
There are notable increases to the required $\S0$ {in} high statistics
due to $\dB$ uncertainties, and the impact is larger 
in $\mathscr{L}_{C}$ than in $\mathscr{L}_{CE}$. 
A $( \dB / B ) {{=}} 10\%$ uncertainty will give rise
to increase in $\S0$ by 45\% and 17\% at $\B0 {{=}} 100$ within RoI 
for counting-only and counting-plus-energy analysis, respectively. 
The availability of the additional energy measurements makes the
evaluation of $\S0$ more robust and less vulnerable to background
uncertainties. 

We note that $\dB$ depends on
the knowledge on $B$ from the auxiliary data prior to
the experiments.
In practice, 
with improving data quality and increasing data size
during the experiments,
$\dB$ can be expected to be further reduced.






\section{Case Study: Neutrinoless Double Beta Decay}
\label{sect::0nubb}


\begin{figure*}
{\bf (a)} \hspace{8cm} {\bf (b)}\\
\includegraphics[width=8.2cm]{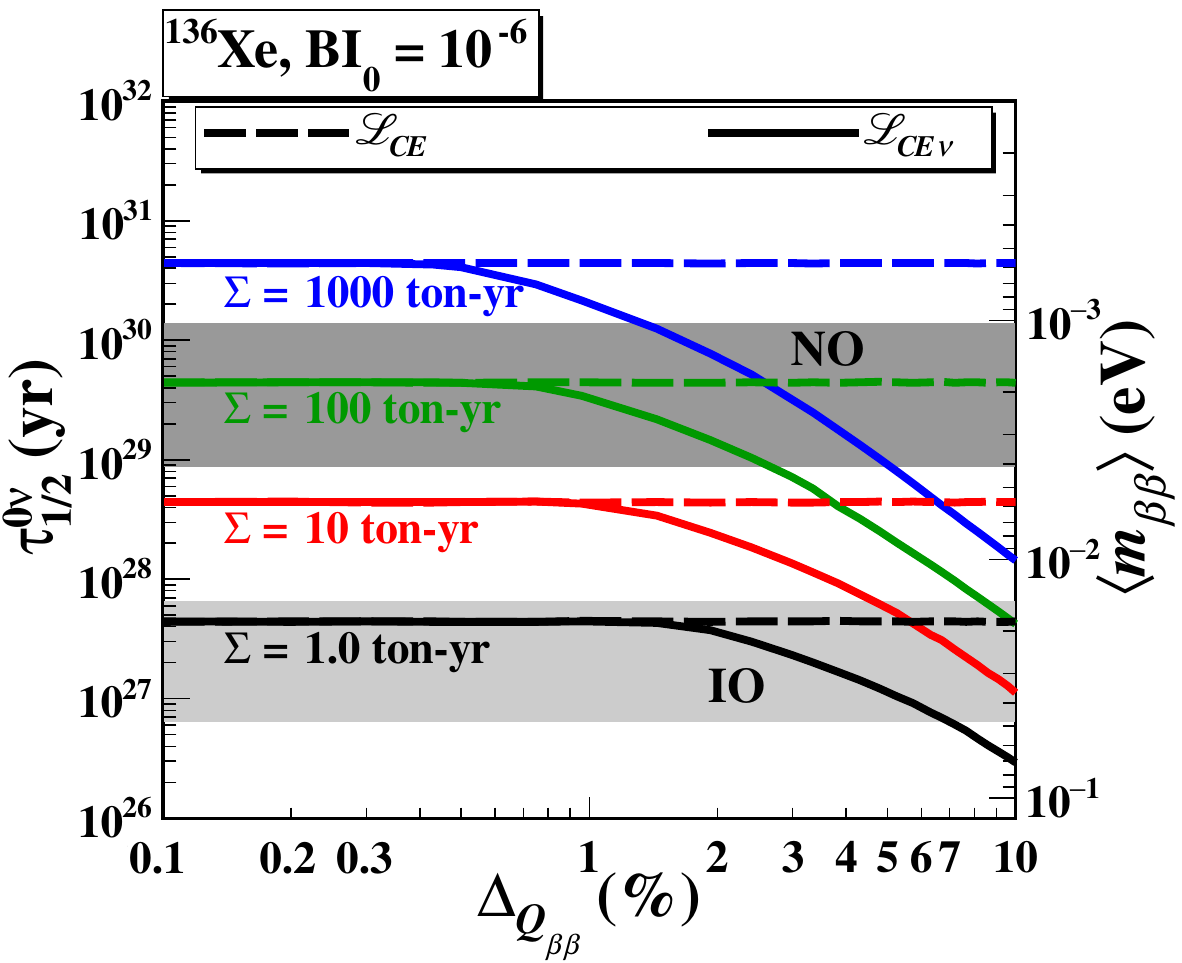}
\includegraphics[width=8.2cm]{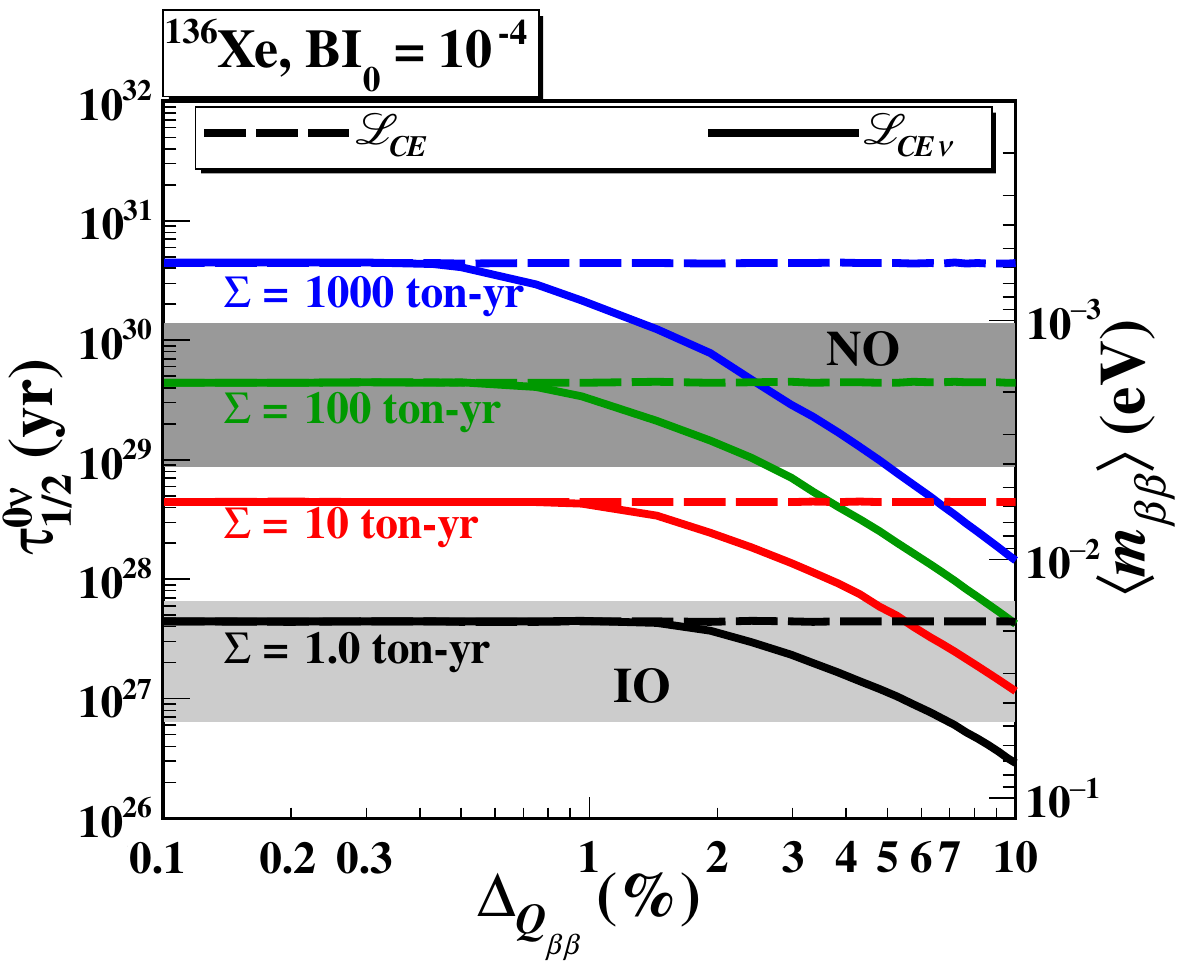}\\
{\bf (c)} \hspace{8cm} {\bf (d)}\\
\includegraphics[width=8.2cm]{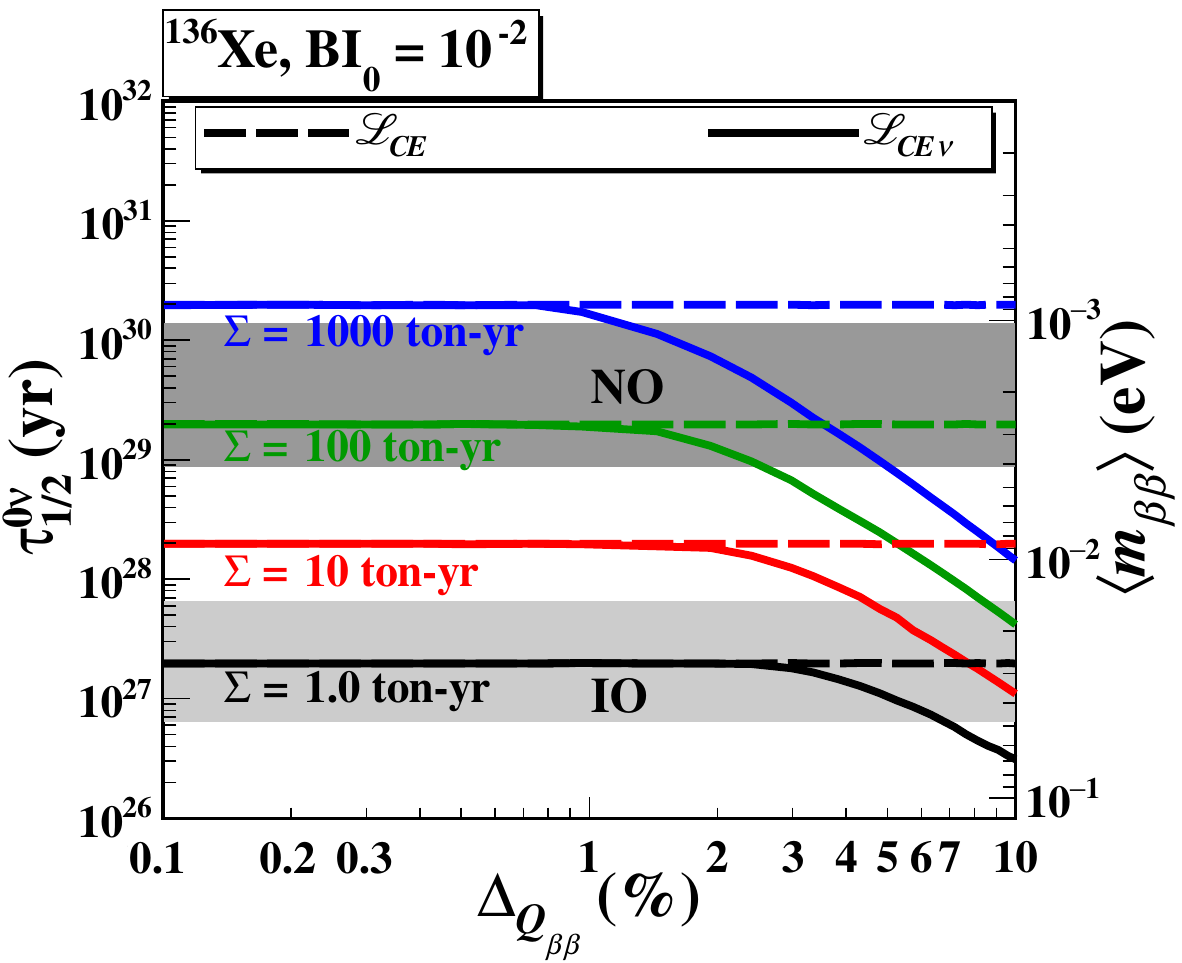}
\includegraphics[width=8.2cm]{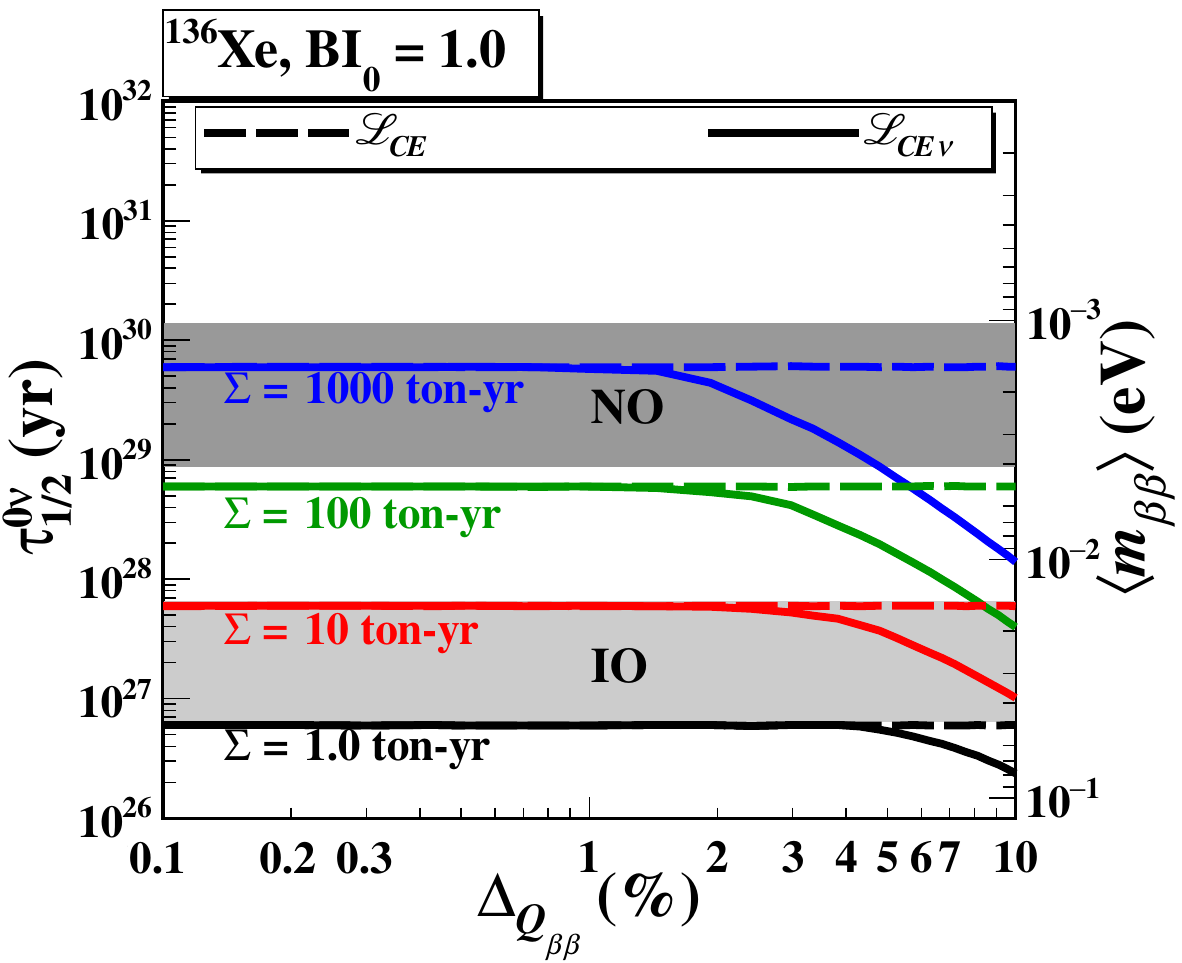}
\caption{
Combined background LLR analysis for $\xe136$
in $( \DeltaQbb , \thalf0nu )$ space at different contours
{
$\Sigma {=} 1,10,100,1000{~}\tyr$
}
taking $\BI0$= (a) $10^{-6}$,
(b) $10^{-4}$, (c) $10^{-2}$,
and (d) 1 $\BIunit$
under the specific case where uncertainties in the
expected ambient background are negligible, or $( \dB/B ) {=} 0\%$.
Case (a) is, in particular, effectively
the zero ambient background condition.
Predicted $\mbb$ ranges for neutrino mass
IO and NO~\cite{Esteban:2020cvm,Huang:2022lub},
following the matrix elements models prescribed in
Ref.~\cite{TEXONO:PRD:2020} are superimposed.
Scenarios with $\2nubb$ background switched off
are displayed as dotted lines to illustrate individual contributions
from both background components.
The $\2nubb$ process is the leading background {for} 
increasing $\DeltaQbb$ beyond the divergent points.
}
\label{fig::combined-bkg-BI}
\end{figure*}


A case study was performed to make
sensitivity projections on future $\0nubb$ experiments
with profile likelihood, similar to previous
work in Ref.~\cite{Adhikari_2022}. 
This study serves to illustrate how
the formulation and algorithms developed in this work
can be applied in practice.
A particular isotope and theoretical model 
are selected as example.
Detailed comparisons taken into account the variety
of target isotopes, experimental design specifications, theoretical modeling
and practical resource-effectiveness
are issues beyond the theme and scope of this work.


The process $\0nubb$~\cite{0nubb-review,RevModPhys.95.025002}
is a lepton-number
violating process involving the decays of isotope $\Abb (N,Z)$
to two electrons:
\begin{equation}
^N_Z \Abb ~ \rightarrow ~
_{Z+2}^{N-2}A ~ + ~ 2 e^-  ~~.
\label{eq::0nubb}
\end{equation}
Experimental signature is a {monoenergetic} energy peak
at the decay Q-value ($\Qbb$).
The FWHM of the $\0nubb$-peak is denoted by $\DeltaQbb$ in \%.


The decay half-life $\thalf0nu$ 
can be derived from measurements via:
\begin{equation}
{  \thalf0nu }  ~ = ~
 { {\rm ln~2} }   \cdot
\left[ \frac{N_A}{ ( N + Z ) } \right] \cdot 
\left[ \frac{ \Sigma }{S_{obs} / \effRoI } \right] ~ ,
\label{eq::roi}
\end{equation}
where $N_A$ is the Avogadro Number,
$\Sigma$ denotes
the combined exposure typically 
expressed in units of ton-year~($\tyr$),
and $S_{obs}$ is the observed strength of
the $\0nubb$ peak. For simplicity,
we take the ideal case where both isotopic abundance and 
experimental signal efficiency are 100\%. 
The realistic exposure 
relative to the ideal one 
can be evaluated by corrections 
on these two parameters~\cite{TEXONO:PRD:2020}. 

The measurable is related to neutrino masses via:
\begin{equation}
\left[ \frac{1}{ \thalf0nu } \right] ~ = ~
G^{0 \nu} ~ g_A^4 ~ \ME2 ~ \left| \frac{\mbb}{m_e} \right| ^2  ~ ,
\label{eq::thalf0nu}
\end{equation}
where
$m_e$ is the electron mass,
$g_A$ is the effective axial vector coupling~\cite{nubb-gA:1,nubb-gA:2},
$G^{0 \nu}$ is a known phase space
factor~\cite{phasespace} due to kinematics,
$| M^{0 \nu} |$ is the nuclear physics
matrix element~\cite{dbdme-review}, while
$\mbb$ is the effective Majorana neutrino mass.
To connect $| M^{0 \nu} |$ with $\mbb$,
we adopt the model of Ref.~\cite{robertson} 
which observed that $\left[ \ME2 \cdot G^{0 \nu} \right]$ 
can be {approximated by} a constant at fixed $\mbb$ independent
of {the $\0nubb$ candidate isotopes}.
Measurements in $\thalf0nu$ can then be 
translated to sensitivities in $\mbb$
and be compared to the predicted ranges of  
neutrino mass Inverted and Normal Ordering (IO and NO)~\cite{Esteban:2020cvm,Huang:2022lub}.


Two background channels are considered:
(i) ambient background which is assumed to be constant at $\Qbb$, and
(ii) background due to two-neutrino double beta decay ($\2nubb$) which
leaks into the $\0nubb$ peaks due to {non-zero} energy resolution
of $\DeltaQbb$.
{
Other background such as cosmogenic-induced events and solar neutrino interactions can be
incorporated in future research, by expanding the constant ambient background conditions
to include additional spectral components with energy dependence.
}

Following conventions~\cite{mbbprob:3,LEGEND:2021_Ge,Zhao_2017},
the ambient background is parametrized by the
``Background Index'' ($\BI0$) defined as:
\begin{equation}
\BI0 ~  \equiv  ~
 \frac{\B0 ( \DeltaQbb ) }{ \Sigma }
\label{eq::BI}
\end{equation}
{which is the background in the FWHM energy range $\DeltaQbb$
  around $\Qbb$ per ton-year of exposure},
with dimension [$\BIunit$].
Background levels expressed in $\BI0$ 
are universally applicable to
{comparing} sensitivities of different
$\0nubb$-experiments on a variety of {the 0$\nu\beta\beta$ candidate isotope}. 


\begin{figure*}
{\bf (a)} \hspace{8cm} {\bf (b)}\\
\includegraphics[width=8.2cm]{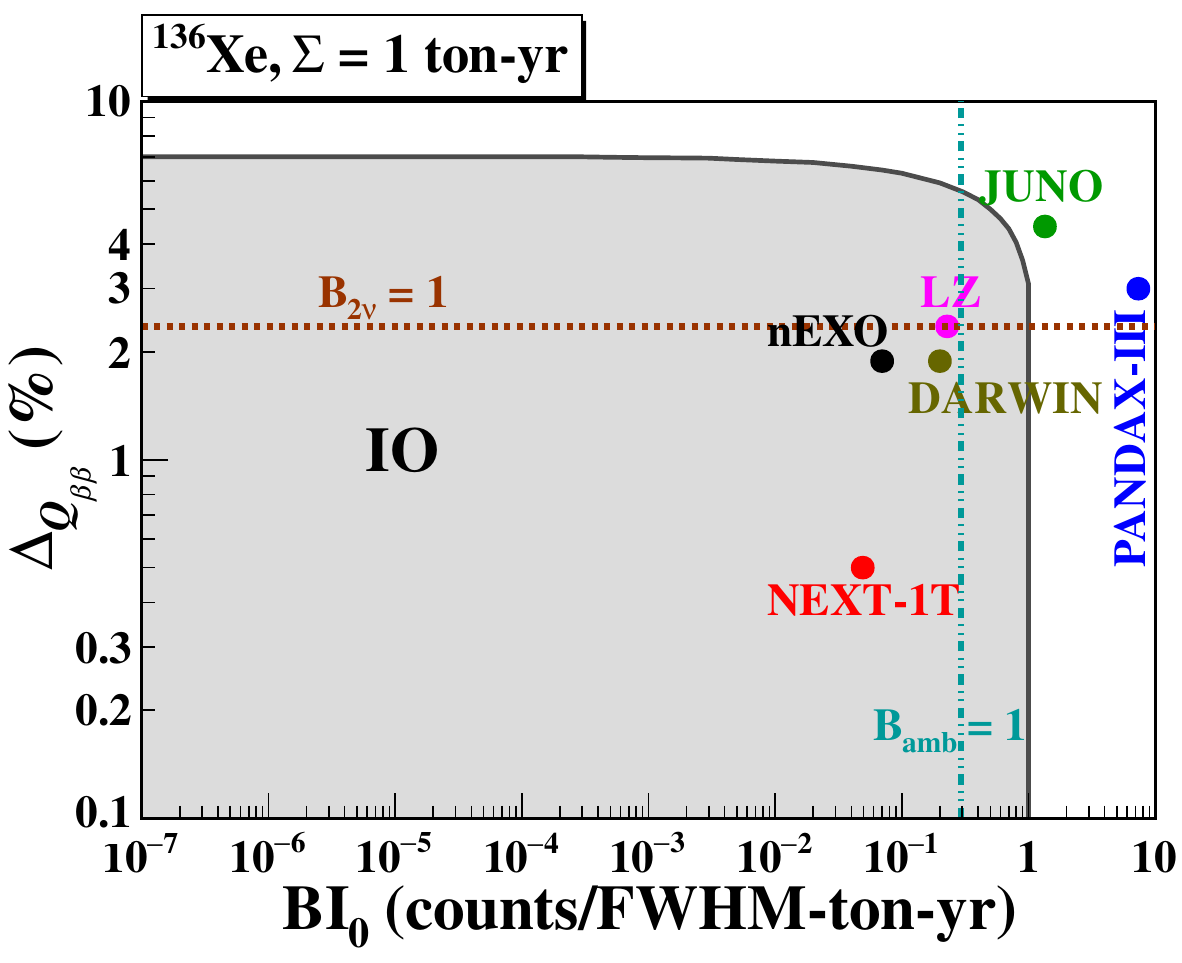}
\includegraphics[width=8.2cm]{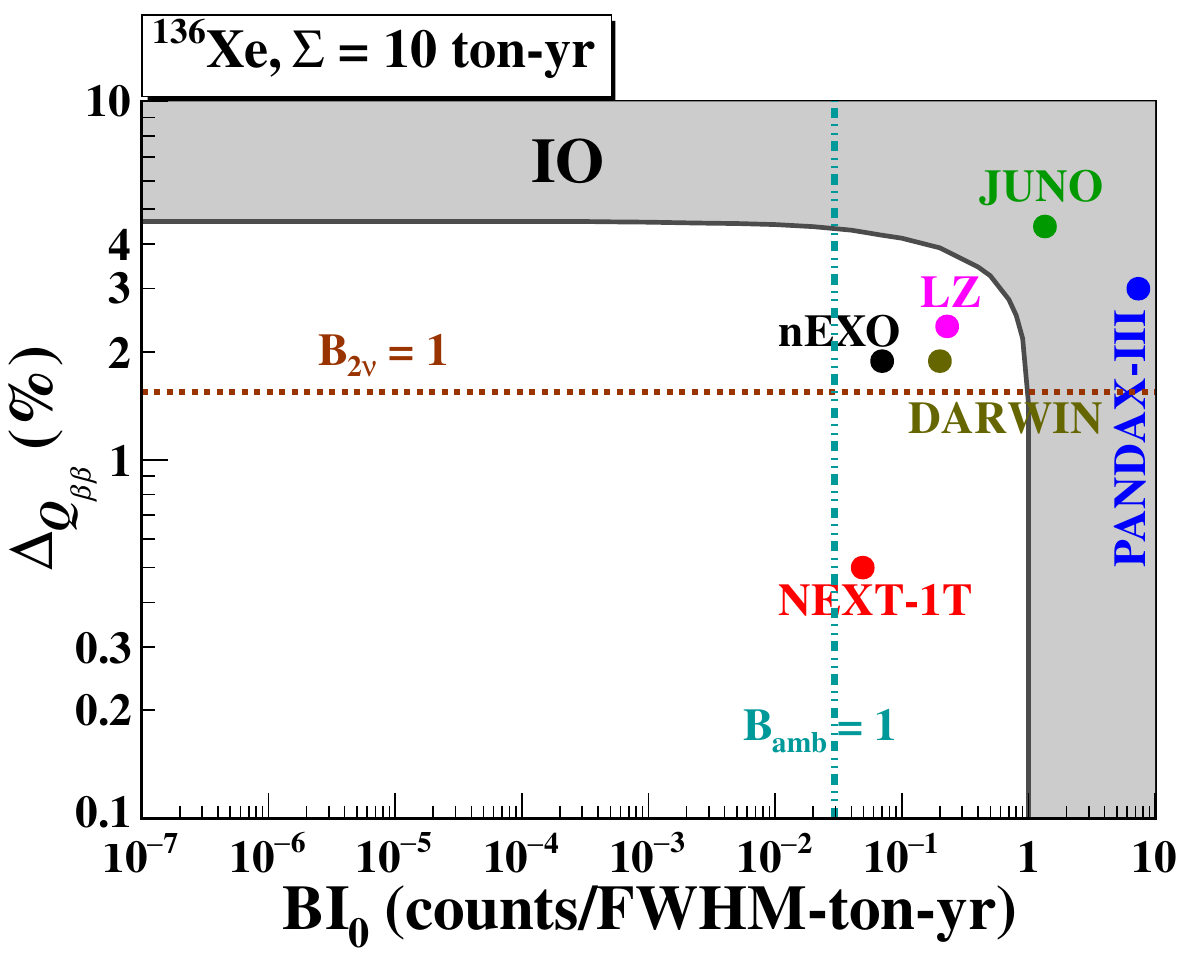}\\
{\bf (c)} \hspace{8cm} {\bf (d)}\\
\includegraphics[width=8.2cm]{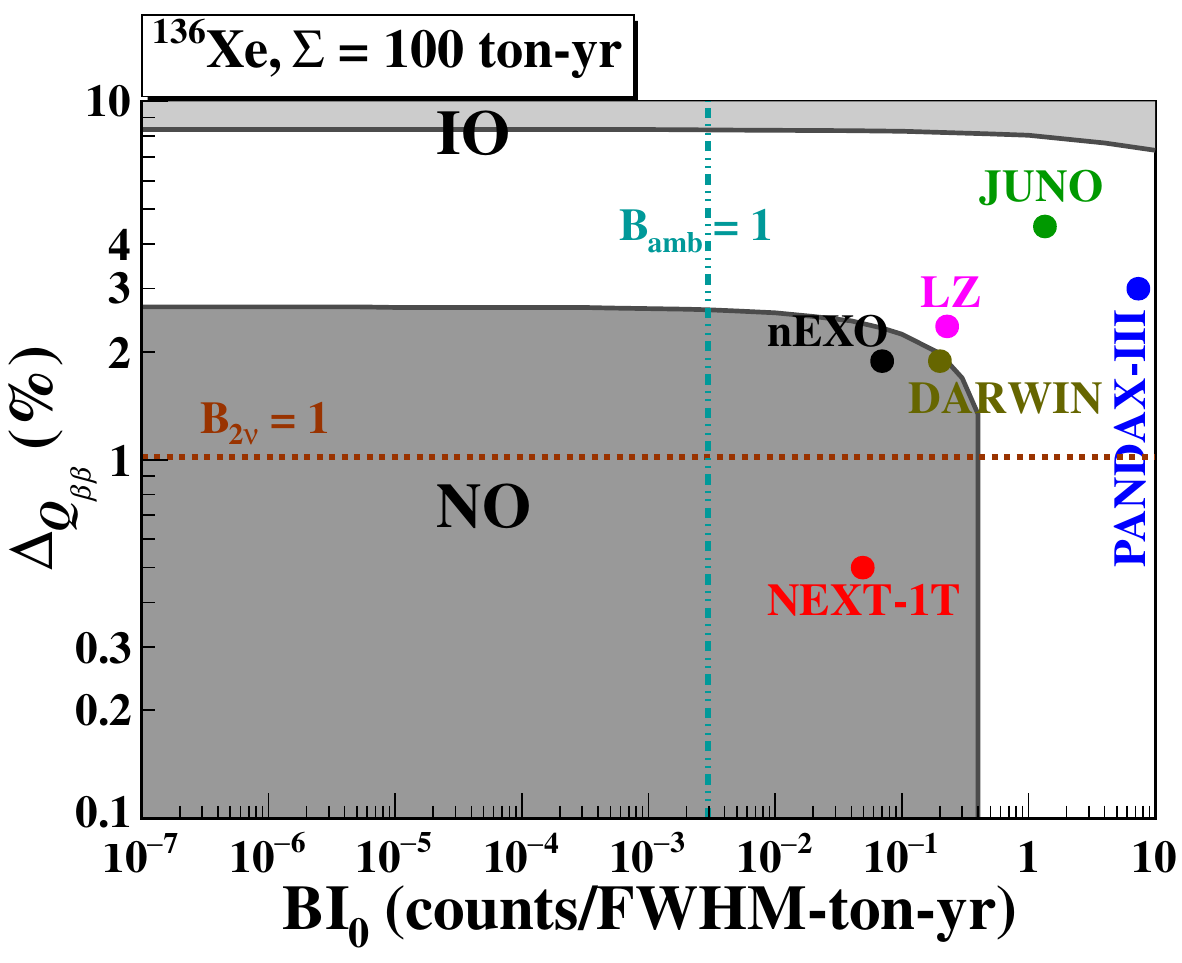}
\includegraphics[width=8.2cm]{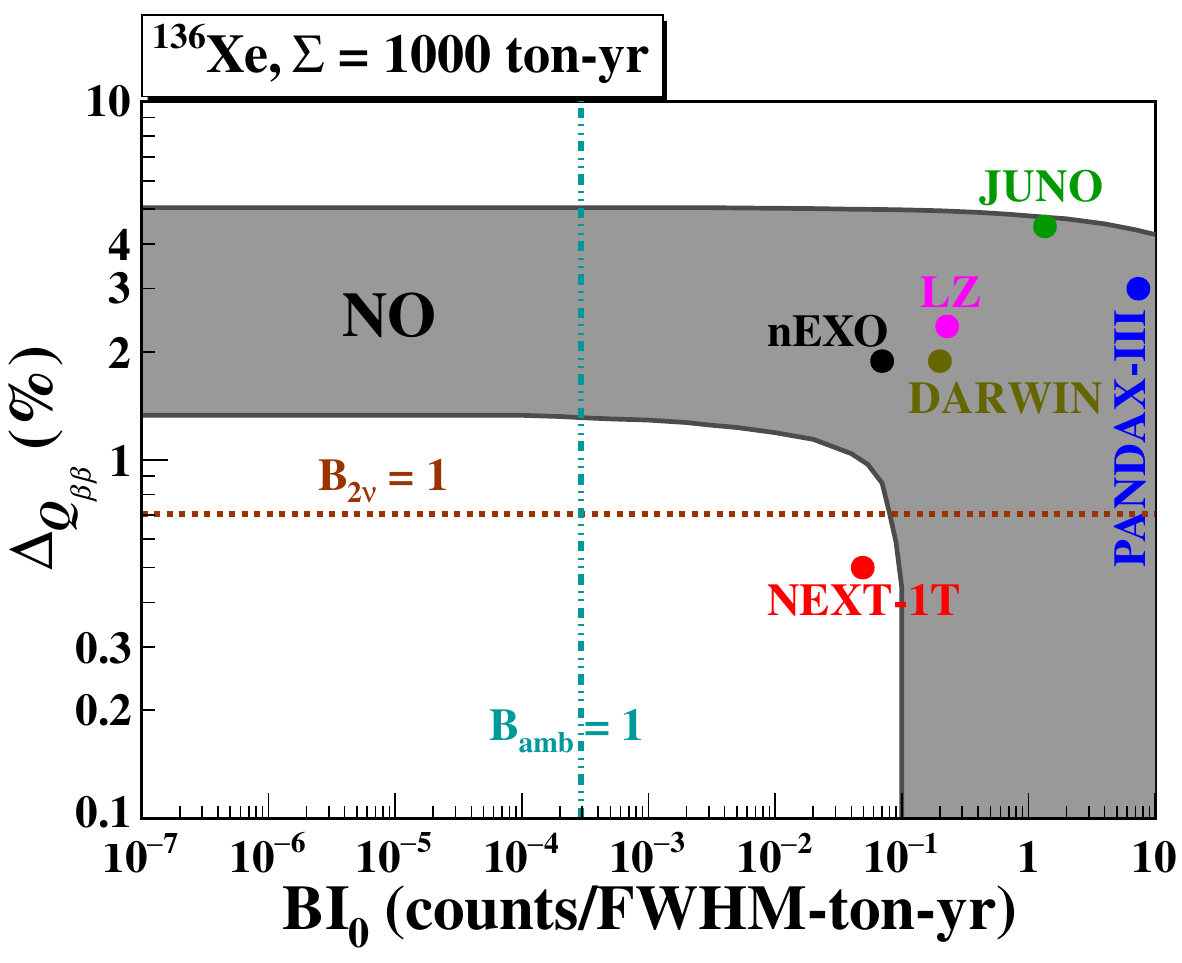}
\caption{
Requirements in $( \BI0 , \DeltaQbb )$ space
for $\0nubb$ experiments with $\xe136$
to achieve $\P503s$, for $\Sigma$ at (a) 1, (b) 10, (c) 100 and (d) 1000 ton-yr,
under the specific case where uncertainties in the
expected ambient background are negligible, or $( \dB/B ) {=} 0\%$.
Detector performance parameters in $( \BI0 , \DeltaQbb )$ for the coming generation of
$\xe136$-projects~\cite{Adhikari_2022,NEXT:2020amj,Han_2020,PhysRevC.102.014602,DARWIN:2020jme}
are superimposed.
The ${\rm B}_{2 \nu} {=} 1$ and ${\rm B}_{\rm amb} {=} 1$
contours correspond to, respectively, where the first 
$\2nubb$ and ambient background event would appear
within ${\rm RoI} {=} \Qbb {\pm} 4 \sigmaE0$.
}
\label{fig::DeltaQbbVsBI0}
\end{figure*}

\begin{figure}
\includegraphics[width=8.2cm]{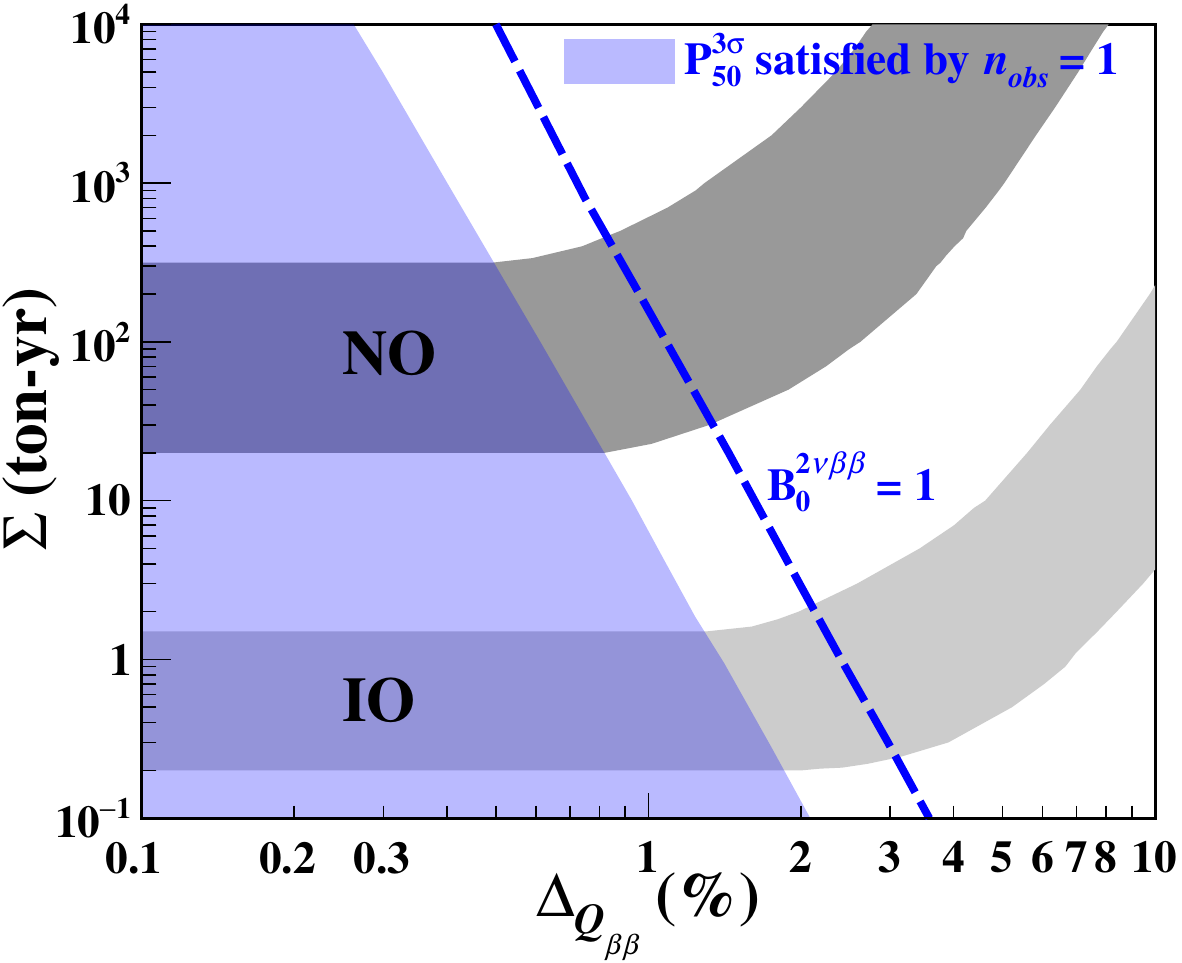}
\caption{ 
The conditions, represented by the white region,
under which the irreducible $\2nubb$ background for $\xe136$ 
limits the $\0nubb$ sensitivities in the zero ambient background scenario.
The blue shaded region corresponds to parameter space 
where one observed event can constitute a positive signal under $\P503s$.
The blue dotted line depicts the case where 
one $\2nubb$ event can be observed on average.
The bands for IO and NO are superimposed.
}
\label{fig::2nubbwall}
\end{figure}

{The input parameters specific to the $\0nubb$
candidate isotope chosen for this study, $\xe136$,
are} $\Qbb {=} 2.458~{\rm MeV}$ and
$\t2bb {=} 2.2 {\times} 10^{21}~{\rm yr}$~\cite{PhysRevC.89.015502,PhysRevC.105.055501,PhysRevC.86.021601}. 
Signal events with strength $\S0$ 
with Gaussian energy distribution
at mean $\Qbb$ and FWHM $\DeltaQbb$
are simulated, superimposed by both 
background channels.
Multiple simulated data sets {for} different 
$( \B0, \S0 )$ are produced.

{The ambient background is assumed to be energy-independent.}
The $\2nubb$ background spectrum with
the parametrization of Ref.~\cite{spec_2nu} is adopted.
The measured spectrum is derived via
Gaussian smearing with width characterized by 
detector resolution $\DeltaQbb$.
The likelihood with expected $\2nubb$ background 
and uncertainties of 
$\dB$ (${=} \sqrt{{\tau}B}/\tau$)
can be written as
\begin{eqnarray}
\label{eq::extended_likelihood_with_2nu}
\mathscr{L}_{CEB \nu} & ~ \equiv ~ & \mathscr{L}( S,B | \mathbb{E} )  \\
& =  & \frac{e^{-( B + \nu + S )}( B {+} \nu {+} S )^{N}}{N!}
\frac{e^{-{\tau} B  }( {\tau}B )^{n_{0}}}{n_{0}!} ~~~~ \nonumber  \\
& \times & \prod_{i=1}^{N}
\left[ \frac{ B  \cdot {f_{B}(E_{i})} {+} {\nu}\cdot f_{2\nu}(E_{i}) {+} S  
\cdot f_{S}(E_{i})}{( B {+} \nu {+} S )} \right] ~ , ~~~~ \nonumber
\end{eqnarray}
where $\nu$ is the expected count of $\2nubb$ in RoI,
and $f_{2\nu}(E)$ is the $\2nubb$ spectrum 
normalized with 
$ \int_{\rm{RoI}}{f_{2\nu}(E)dE} {=} 1 $.
{
We first take the asymptotic case of
$( \dB/B ) {\simeq} 0\%$ with the $\tau B$-term suppressed.
The likelihood of Eq.~\ref{eq::extended_likelihood_with_2nu} is simplified to
\begin{eqnarray}
\label{eq::extended_likelihood_with_2nu_No-sigmaB}
\mathscr{L}_{CE \nu} 
&  ~ =  ~  & \frac{e^{-( B + \nu + S )}( B {+} \nu {+} S )^{N}}{N!} \\
& \times & \prod_{i=1}^{N}
\left[ \frac{ B  \cdot f_{B}(E_{i}) {+} {\nu}\cdot f_{2\nu}(E_{i}) {+} S  
\cdot f_{S}(E_{i})}{( B {+} \nu {+} S )} \right] ~ . ~~~~ \nonumber
\end{eqnarray}
Uncertainties of $2\nu\beta\beta$ background rates and spectra 
are also negligible in this analysis.
}

The LLR analyses are applied to cases 
with and without $\2nubb$ background described
by likelihood functions of, respectively, 
$\mathscr{L}_{CEB \nu}$ in Eq.~\ref{eq::extended_likelihood_with_2nu}
and
$\mathscr{L}_{CE}$ in Eq.~\ref{eq::extended_likelihood}.
Distributions of $\q0$ following Eq.~\ref{eq::q0} 
for $\nll-Pq0H0$ and $\alt-Pq0H1$ in
low and high statistics scenario,
similar to those of Figures~\ref{fig::MLR-B0}a\&b, 
are derived.
The $F(E_{i}|S,B)$ in Asimov data set 
of Eq.~\ref{eq::binned_expected_value} 
is expanded to
\begin{equation}
F(E_{i}|S,B) {=} \left[ B  \cdot {f_{B}(E_{i})} 
{+} \nu \cdot f_{2\nu}(E_{i}) 
{+} S \cdot f_{S}(E_{i}) \right]  w ( E_i )
\label{eq::binned_expected_value_2nu}
\end{equation}
with an additional $[ \nu \cdot f_{2\nu}(E_{i}) ]$ factor.


The $\thalf0nu$ versus $\DeltaQbb$  
{
at different contours of $\Sigma {=} 1,10,100,1000{~}\tyr$
} 
scanning over $\BI0 {=} {10^{-6}, 10^{-4}, 10^{-2}, 1} {~} \BIunit$
are depicted in Figures~\ref{fig::combined-bkg-BI}a,b,c\&d,
{superimposed on} the predicted ranges of IO and NO~\cite{Esteban:2020cvm,Huang:2022lub}.
The divergent points between the solid and dotted lines depend on $\Sigma$ and $\BI0$.
They denote the $\Delta$-values above which 
the irreducible $\2nubb$ background would dominate.
In particular, the low $\BI0 {=} 10^{\mbox{-}6}$ scenario 
of Figure~\ref{fig::combined-bkg-BI}a
corresponds to where the ambient background can be neglected.

The allowed regions to achieve $\P503s$ in
$ ( \DeltaQbb , \BI0 )$ space for 
$\Sigma {=}  1, 10, 100, 1000 ~ {\rm \tyr}$
are depicted in Figures~\ref{fig::DeltaQbbVsBI0}a,b,c\&d,
in which the performance specifications in $( \BI0 , \DeltaQbb )$ for the coming generation of 
$\xe136$-projects~\cite{Adhikari_2022,NEXT:2020amj,Han_2020,PhysRevC.102.014602,DARWIN:2020jme} 
are superimposed. 
{For} fixed $\Sigma$,
ambient and $\2nubb$ background depend
only on $\BI0$ and $\DeltaQbb$, respectively. The contours of 
${\rm B}_{2 \nu} {=} 1$ and ${\rm B}_{\rm amb} {=} 1$ 
within ${\rm RoI} {=} \Qbb {\pm} 4 \sigmaE0$ are marked.

  { 
While the numerical results are derived from $\xe136$ under the
assumptions stated,
some general and notable features related to the sensitivity projections 
for future $\0nubb$ projects can be observed:
}
\begin{enumerate}
  { 
\item
Following Figure~\ref{fig::comparison},
counting-only analysis can lead to sensitivity projections which deviate 
by ${>} 6\%$ from those of complete LLR analysis with energy information included.
The discrepancies can be as large as 20-30\%
for $ \BI0 \Sigma {<} 10^{\mbox{-}2} $. 
\item
The point at which the solid and dotted lines converge signifies 
the transition on which of the two background modes are dominant
$-$ the ambient and $\2nubb$ background dominate the sensitivities
at $\DeltaQbb$ values lower and higher than the transition point, respectively.
\item 
Effects of non-zero $( \dB/B )$: 
At parameter space in Figure~\ref{fig::combined-bkg-BI} 
where $\2nubb$ background dominates, there are no effects to the sensitivities.
When ambient background is the leading channel, 
the relative drop of sensitivities (equivalently, increase in required $\Sigma$)
can be read off directly from Figure~\ref{fig::deltaB}c.
\item
The low-$\DeltaQbb$ regime in Figure~\ref{fig::combined-bkg-BI}a 
for $\BI0 {=} 10^{\mbox{-}6}$ is effectively the zero ambient background condition.
The blue shaded region in Figure~\ref{fig::2nubbwall} 
is where background due to $\2nubb$ is also negligible
such that one observed event within RoI 
will constitute a positive signature under $\P503s$. 
The required experimental specifications are
$\DeltaQbb {<} 1.3\%$ and $\Sigma {>} 1.5$ ton-year for IO, 
and $\DeltaQbb {<} 0.5\%$ and $\Sigma {>} 315$ ton-year for NO.
The white region is where the irreducible $\2nubb$ background 
limits the $\0nubb$ sensitivities. 
The blue dotted line depicts the case where one $\2nubb$ background event 
can be observed on average.
}
\item
The relatively high background levels of 
$\BI0 {=} 1$ in Figure~\ref{fig::combined-bkg-BI}d 
corresponds to those achieved 
in the current generation of experiments~\cite{PhysRevLett.125.252502}.
The $\2nubb$ background is only of minor impact except for
$\DeltaQbb$ larger than a few\% where the solid and dotted lines diverge.
Exposures of $\Sigma {{=}} 10 ~ \tyr$
and 100~$\tyr$ are required to cover IO from experiments
with $\DeltaQbb {<}$1.4\% and 8.0\%, respectively.
In addition, probing the entire NO region
is not possible even with $\Sigma {\sim} 1000 ~ \tyr$ 
for experiments with {$\DeltaQbb {=} 0.12\%$~\cite{PhysRevLett.120.132502},
the best resolution achieved to-date with $\ge76$}.

\item
It can be inferred from Figure~\ref{fig::DeltaQbbVsBI0}b that
the experimental specifications for the coming generation of projects 
could cover IO at $\Sigma {>} 10 ~ {\rm \tyr}$. 
However, following Figure~\ref{fig::DeltaQbbVsBI0}d,
this would be insufficient to probe NO.
Covering NO entirely would require
$\Sigma {\simeq} 1000~\tyr$ at
$\DeltaQbb {\lesssim} 1\%$ together with 
$\BI0$ at ${\lesssim} 0.1$.
\item
Future $\0nubb$ projects to probe IO and NO 
would necessarily have $ ( \BI0 \cdot \Sigma ) {<} 1$ 
with multiple ton-year exposure of enriched isotopes.
A mis-estimation of the sensitivity reach by a few-\% 
already implies non-optimal use of substantial resources.
It follows from Figure~\ref{fig::comparison}b that 
counting-only analysis with complete Poisson or continuous approximations 
are no longer adequate. Energy information has to be incorporated in 
the evaluation of the sensitivity projections to 
provide the best input for the assessment of cost-effectiveness.
\end{enumerate}


\section{Summary and Prospects}

We develop in this work the statistical methods to define required signal strength 
to establish a positive effect in 
an experiment with known background and uncertainties $-$ before it is performed.
It expands from our earlier counting-only analysis~\cite{TEXONO:PRD:2020} 
to incorporate constraints from additional measurements.  

Two expected features are quantified on
the required signal strength to establish positive effects.
Firstly, in counting-only experiments,
the strength can be derived correctly with complete Poisson analysis,
and the continuous approximation would underestimate the values.
Furthermore, incorporating continuous variables as additional constraints  
would reduce the required signal strength relative to that derived with
counting-only analysis.

The procedures are applied to $\0nubb$ experiments on one 
isotope $\xe136$ under realistic parameters
as illustrations on how they are used in practice.
The theme of our future research would be to adapt these tools to 
perform systematic studies on the sensitivity dependence of $\0nubb$ projects 
to experimental choice of target isotopes, detector resolution and planned exposure.

\section{Acknowledgment}

This work is supported by 
the Academia Sinica Principal Investigator Award 
AS-IA-106-M02,
contracts 106-2923-M-001-006-MY5,
107-2119-M-001-028-MY3 and
110-2112-M-001-029-MY3,
from the Ministry of Science and Technology, Taiwan,
and
2021/TG2.1 from 
the National Center of Theoretical Sciences, Taiwan.
M.K.~Singh and H.B.~Li make equal contributions to this work.


\bibliography{dcpbkg-ref}

\end{document}